\documentclass[10pt,prb,twocolumn,aps,superscriptaddress, longbibliography,nofootinbib]{revtex4-2}
\usepackage{graphicx}
\usepackage{amsmath,xcolor}
\usepackage[T1]{fontenc}
\usepackage{enumitem}
\usepackage{multirow}
\usepackage{quantikz}
\usepackage[colorlinks=true,citecolor=blue]{hyperref}
\usepackage{csquotes}
\usepackage{lipsum}
\usepackage{mwe}
\usepackage{textpos}
\usepackage{tikz-3dplot}
\usepackage[normalem]{ulem}
\makeatletter
\renewcommand\@makefnmark{\hbox{\textsuperscript{\normalfont\@thefnmark}}}
\makeatother

\begin{document}
    \title{Singlet-only always-on gapless exchange (SAGE) spin qubits: Charge noise effects and two-qubit gates}

    \author{Nathan L. Foulk$^{\dagger,*}$}
    \affiliation{Condensed Matter Theory Center and Joint Quantum Institute, Department of Physics, University of Maryland, College Park, Maryland 20742-4111 USA}

    \author{Katharina Laubscher$^{\dagger}$}
    \affiliation{Condensed Matter Theory Center and Joint Quantum Institute, Department of Physics, University of Maryland, College Park, Maryland 20742-4111 USA}

    \author{Silas Hoffman$^{\ddagger}$}
    \affiliation{Condensed Matter Theory Center and Joint Quantum Institute, Department of Physics, University of Maryland, College Park, Maryland 20742-4111 USA}
    \affiliation{Laboratory for Physical Sciences, 8050 Greenmead Drive, College Park, Maryland 20740, USA}

    \author{Sankar Das Sarma}
    \affiliation{Condensed Matter Theory Center and Joint Quantum Institute, Department of Physics, University of Maryland, College Park, Maryland 20742-4111 USA}

\begin{abstract}
Singlet-only always-on gapless exchange (SAGE) spin qubits are an alternative type of exchange-only (EO) qubits that encode a single qubit in the spins of four electrons located in four tunnel-coupled quantum dots. While conventional EO qubits are susceptible to local magnetic field gradients caused by local nuclear environments and $g$-factor variations, the SAGE qubit subspace is inherently protected from magnetic-gradient-induced Pauli errors by virtue of the singlet-only encoding, which is invariant under magnetic field gradients, and the always-on exchange couplings, which provide energetic leakage protection. However, the always-on operation simultaneously increases the qubit's sensitivity to charge noise. Here, starting from a Hubbard model describing the underlying electronic structure of the coupled quantum dots, we characterize the performance of SAGE qubits in the presence of $1/f$ charge noise that induces fluctuations in both the dot chemical potentials and the interdot tunnel couplings. We calculate SAGE idle coherence times and show that realistic CPMG-like pulse sequences can be used to significantly extend SAGE single-qubit coherence times for experimentally relevant charge noise strengths. We likewise study the fidelity of SAGE two-qubit gates in the presence of charge and magnetic noise and again propose a simple refocusing strategy to mitigate the noise, while increased ramp times of the entangling pulse suppress leakage into noncomputational states.
\end{abstract}

\maketitle

\begingroup
\renewcommand\thefootnote{\fnsymbol{footnote}}
\footnotetext[1]{Nate.Foulk@jhuapl.edu; Johns Hopkins Applied Physics Laboratory, Laurel, Maryland 20723.}
\footnotetext[2]{These authors contributed equally to this work.}
\footnotetext[3]{Present address: Microsoft Quantum.}
\endgroup

\section{Introduction}
\label{sec:intro}

Exchange-only (EO) spin qubits have recently attracted significant experimental interest~\cite{andrews_quantifying_2019,kerckhoff_magnetic_2021,ha_flexible_2022,weinstein_universal_2023,heinz_fast_2024,acuna_coherent_2024,Madzik2025,Broz2025,Broz2026} as a potential alternative to conventional Loss-DiVincenzo (LD) spin qubits~\cite{loss_quantum_1998}. Here, quantum information is encoded in collective spin states that can be manipulated using only exchange interactions, enabling arbitrary single-qubit rotations via baseband control of exchange couplings between neighboring dots. These baseband pulses are implemented through time-dependent bias voltages, which closely matches the natural capabilities of integrated electronics. By contrast, microwave-controlled qubits require more complex cryogenic circuitry, increasing power dissipation and associated heating~\cite{bartee2025spinqubit}, which will degrade the performance of a quantum processor, particularly as qubit counts increase toward fault-tolerant regimes. Moreover, the use of local exchange control helps avoid challenges associated with global drives, such as rf crosstalk and the ac Stark effect~\cite{Undseth2023,Lawrie2023}, and eliminates the need for micromagnets to induce synthetic spin-orbit interaction.

Despite these advantages, conventional EO qubits---which are encoded using three spins across three quantum dots~\cite{divincenzo_universal_2000}---also come with notable challenges. While immune to global magnetic field fluctuations, conventional EO qubits are sensitive to local magnetic field gradients arising from nuclear spin environments (or $g$-factor variations between dots), which induce both coherent errors within the qubit subspace and leakage into noncomputational states. Moreover, implementing high-fidelity EO two-qubit gates typically requires complex pulse sequences to suppress leakage out of the computational subspace~\cite{fong_universal_2011}. These problems have hindered experimental implementations of EO qubits in spite of their obvious advantages over LD-type qubits without any necessity for engineering artificial spin-orbit coupling through the use of micromagnets for single-qubit operations.

To overcome the limitations associated with magnetic-gradient-induced errors, Ref.~\cite{sala_exchange-only_2017} proposed the exchange-only singlet-only (XOSO) qubit. Here, information is encoded in the total singlet subspace of four electrons in four quantum dots, which renders the computational subspace inherently immune to coherent errors caused by local magnetic field gradients~\cite{bacon_robustness_1999,bacon_universal_2000,bacon_encoded_2001,kempe_encoded_2001,kempe_theory_2001,bacon_decoherence_2003,sala_exchange-only_2017,sala_highly_2020}. At the same time, leakage to noncomputational states can be energetically suppressed by keeping the exchange couplings always-on~\cite{weinstein_energetic_2005}. In Ref.~\cite{Foulk2025}, a gapless version of such a singlet-only always-on qubit---the singlet-only always-on gapless exchange (SAGE) qubit---was characterized, where baseband pulses are used to lower exchange couplings  away from a finite-exchange idle point, thereby restoring one of the control advantages that initially motivated the development of EO qubits. SAGE qubits exhibit significantly longer idle coherence times and higher single-qubit gate fidelities than conventional EO qubits in regimes where magnetic noise is the dominant source of errors and, moreover, allow for a fully entangling two-qubit operation using a single interqubit exchange pulse~\cite{Foulk2025} rather than the elaborate multi-step sequences required in conventional EO implementations. These features make SAGE qubits an attractive alternative to standard EO qubit architectures.

\begin{figure*}[tb]
	\centering
	\includegraphics[width=1\linewidth]{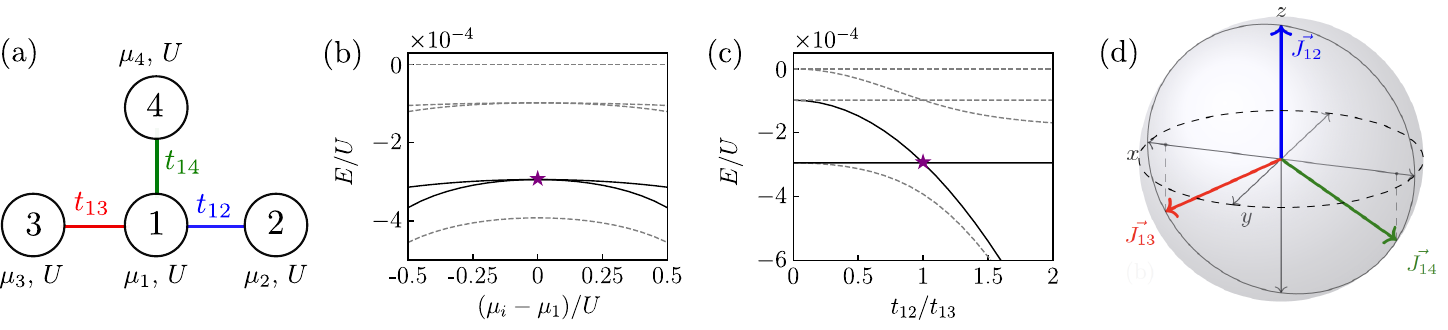}
	\caption{(a) The SAGE qubit is defined using four dots arranged in a T-shape geometry with nearest-neighbor dots tunnel-coupled as indicated by the colored lines. (b,c) Low-energy spectra obtained from Eq.~(\ref{eq:hubbard}) for a system with four electrons and total $S_z=0$ as a function of dot detuning and tunnel coupling, respectively. The solid lines correspond to the computational states, whereas the dashed lines represent leakage states. (b) Energy spectrum as a function of $\mu_i-\mu_1$ (here $i\in\{2,3,4\}$) for $t_{12}=t_{13}=t_{14}=0.007U$, $\mu_i+\mu_1=0$ (symmetric detuning), and $\mu_j=0$ for $j\neq 1,i$. The point $\mu_i=0$ for all $i$ corresponds to a full charge noise sweet spot for SAGE qubits, where the qubit is to first order insensitive to chemical potential fluctuations. (c) Energy spectrum as a function of $t_{12}/t_{13}$. Here $\mu_i=0$ for all $i$ and $t_{13}=t_{14}=0.007U$. The idle point of the SAGE qubit is indicated by a star in both (b) and (c). (d) Bloch sphere representation of the SAGE qubit showing the rotation axis of control for each exchange coupling. For small $t_{ij}\ll U$ the exchange couplings are approximately given by Eq.~(\ref{eq:J}).}
	\label{fig:setup}
\end{figure*}

However, because the exchange interactions must be kept always on in order to energetically suppress leakage errors, the SAGE qubit---like all always-on qubits---exhibits an increased sensitivity to charge noise, which is unavoidably present in realistic semiconductor devices. In particular, fluctuations in the gate voltages as well as uncontrolled charge switching associated with impurities and defects invariably lead to random time-dependent modulations of the exchange couplings, thus limiting SAGE coherence times and gate fidelities despite the qubit's excellent protection against magnetic-gradient-induced noise (which corrects the most serious error source in other EO qubits). In contrast to conventional EO qubits, where the exchange is activated only during gate operations, charge noise strongly affects SAGE qubits even during idle periods. As a result, the overall performance of SAGE qubits will depend almost completely on how effectively charge noise can be controlled and/or mitigated in real experimental systems. We mention here as an aside that charge noise, in some form or other, is the most important decoherence problem for quantum computing in general, limiting the coherence time in all solid-state qubit platforms, not just for semiconductor spin qubits.

Motivated by these considerations, this work presents a detailed study of the effects of $1/f$ charge noise, and potential mitigation strategies, for SAGE qubits. We model each SAGE qubit using a Hubbard-like model describing four electrons in four quantum dots and incorporate (i) $1/f$ charge noise as fluctuations of the gate potentials controlling the individual dot detunings and interdot tunnel couplings and (ii) quasistatic magnetic noise as local fluctuations of the onsite magnetic fields. It is known that the Hubbard model description works well for spin qubits because the exchange coupling essentially derives from the Hubbard coupling in the quantum dots forming the spin qubits~\cite{yang2011generic,yang2011quantum,Dassarma2011}. We show that the performance of SAGE qubits can be significantly improved by implementing dynamical decoupling techniques that effectively cancel out low-frequency charge fluctuations. In particular, we demonstrate that pulse sequences similar to Carr-Purcell (CP)/Carr-Purcell-Meiboom-Gill (CPMG)~\cite{Carr1954Effects,Meiboom1958Modified} can lead to SAGE single-qubit coherence times on the order of hundreds of $\mu$s even for significant charge noise strengths, bringing SAGE qubits on par with conventional spin qubits in charge-noise dominated regimes, while at the same time maintaining their superiority in magnetic-noise dominated regimes. Similarly, dynamical decoupling strategies can also be used to significantly improve the fidelity of SAGE two-qubit gates. Our results indicate that dynamical decoupling offers a practical strategy for mitigating the central weakness of SAGE qubits, positioning them as a highly attractive building block for a scalable all-electrical spin qubit architecture that is relatively free of the leakage that the current EO paradigm is grappling with.

This paper is organized as follows. In Sec.~\ref{sec:model}, we introduce the four-dot Hubbard model used to describe the SAGE qubit, as well as our noise model used to incorporate $1/f$ charge noise and quasistatic magnetic noise into the system. Next, in Sec.~\ref{sec:singlequbit}, we characterize the behavior of individual SAGE qubits in the presence of quasistatic magnetic and $1/f$ charge noise. In particular, we study SAGE idle coherence times with and without dynamical decoupling, showing that pulse sequences of CPMG-type can significantly enhance the $T_2$ times of SAGE qubits. We then extend our analysis to SAGE two-qubit systems in Sec.~\ref{sec:twoqubit}. We discuss the interplay of intrinsic leakage and noise-induced errors and study the fidelity of entangling two-qubit gates---in particular, the controlled-Z (CZ) gate---as a function of various system parameters. Additionally, we propose a simple echo pulse sequence that can help mitigate the effects of charge noise during SAGE two-qubit gates. Finally, we conclude in Sec.~\ref{sec:conclusions}.

\section{Model}
\label{sec:model}

\subsection{Hubbard model and SAGE qubit definition}
\label{subsec:hubbard}

We model the SAGE qubit via a Hubbard model describing four tunnel-coupled dots in a layout as shown in Fig.~\ref{fig:setup}(a),
\begin{equation}
H_0=U\sum_i  n_{i\uparrow}n_{i\downarrow}-\sum_{i,s}\mu_i n_{is}+\sum_{\langle i,j\rangle,s}t_{ij}(c_{is}^\dagger c_{js}+\mathrm{H.c.}).\label{eq:hubbard}
\end{equation}
Here, $i\in\{1,2,3,4\}$ labels the dots, $\langle i,j\rangle$ runs over pairs of nearest-neighbor dots as indicated by the colored lines in Fig.~\ref{fig:setup}(a), $c_{is}^\dagger$ is the creation operator of an electron of spin $s$ on dot $i$, $n_{is}=c_{is}^\dagger c_{is}$, $U$ is the onsite Hubbard interaction, which we assume to be the largest energy scale in the system, $\mu_i$ is the chemical potential on dot $i$, and $t_{ij}$ is the hopping amplitude for an electron hopping between dots $i$ and $j$. Experimentally, the dot chemical potentials $\mu_i$ are controlled by plunger gates, $\mu_i=\alpha V_i$, where $V_i$ is the gate potential of plunger gate $i$ and $\alpha$ is a lever-arm. Similarly, the tunneling amplitude between a pair of dots $i$ and $j$ is controlled by an exchange gate $V_{ij}$, $t_{ij}= \gamma\exp{(V_{ij}/\beta)}$, where we have assumed an exponential dependence with `exchange tunability' $\beta$. Throughout this work, we use $\alpha=0.2$~meV/mV, $\beta\approx 60$~mV, and $\gamma\approx0.45~\mu$eV, which we choose based on recent experimental results~\cite{Madzik2025}.

The SAGE qubit is defined in the (1,1,1,1) charge sector, where each dot is singly occupied. To lowest order in $t_{ij}/U$, Eq.~(\ref{eq:hubbard}) gives rise to an effective Heisenberg Hamiltonian
\begin{equation}
H_\textrm{eff} = \frac{1}{4}\sum_{j=2}^4 J_{1j}\, \boldsymbol{s}_1 \cdot \boldsymbol{s}_j,\label{eq:heisenberg}
\end{equation}
where the exchange couplings are given by
\begin{equation}
J_{ij}\approx 2t_{ij}^2\left(\frac{1}{U+\mu_i-\mu_j}+\frac{1}{U+\mu_j-\mu_i}\right).\label{eq:J}
\end{equation}
Here, $\boldsymbol{s}_i$ denotes the Pauli vector representing the spin on dot $i$. We note that, in Eq.~(\ref{eq:hubbard}), we have assumed for simplicity that the Coulomb interaction is the same on all dots. Similarly, we have neglected cross-capacitance terms $\propto n_in_j$. Both of these assumptions could easily be lifted, which would only lead to a renormalization of the effective exchange interactions given here.

The Heisenberg Hamiltonian preserves the total spin $S$ and the total spin-$z$ projection $S_z$ of the total four-electron state. We further assume that there is a large background magnetic field, such that states that differ in their total spin projection $S_z$ are energetically separated. The SAGE qubit is then defined in the two-dimensional total singlet subspace with $S_z=S=0$. Explicitly, at zero interdot tunneling, the SAGE computational basis states are given by
\begin{gather}
\ket{0} = \ket{S_{12}S_{34}}, \\
\ket{1} = \frac{1}{\sqrt{3}}\Bigl(\ket{T^0_{12}T^0_{34}} - \ket{T^+_{12}T^-_{34}}- \ket{T^-_{12}T^+_{34}}\Bigl),
\end{gather}
where $\ket{S_{ij}}$ refers to a singlet spin state and $\ket{T_{ij}^0}$, $\ket{T_{ij}^+}$, and $\ket{T_{ij}^-}$ refer to triplet spin states with $S_z =\{0,1,-1\}$, respectively, on dots $i$ and $j$. In addition to the qubit states, there are four leakage states with $S_z=0$ and $S=1,2$ that are energetically close to the qubit states, see Figs.~\ref{fig:setup}(b,c) for example energy spectra as a function of dot detuning and interdot tunnel coupling, respectively. All other leakage states with $S_z\neq 0$ are split away by the large background magnetic field such that we neglect them in the following.

Projecting the effective Heisenberg Hamiltonian onto the SAGE basis states, we obtain the effective SAGE qubit Hamiltonian
\begin{equation}
 H_\textrm{q} =  \frac{J_{14}}{4} (\sqrt{3}\sigma_x + \sigma_z)
- \frac{J_{13}}{4}(\sqrt{3}\sigma_x - \sigma_z) - \frac{J_{12}}{2} \sigma_z,\label{eq:qubit}
\end{equation}
where the $\sigma_i$ for $i\in\{ x,y,z\}$ are the standard Pauli matrices acting in the qubit subspace, and the $J_{ij}$ were given in Eq.~(\ref{eq:J}). This effective qubit Hamiltonian was analyzed in detail in Refs.~\cite{sala_exchange-only_2017, Foulk2025}. In particular, the gap between qubit and leakage states is maximized when $J_{12}=J_{13}=J_{14}\equiv J_0$, while the qubit states remain gapless at this point. This point defines the idle point of the SAGE qubit, which is indicated by a star in Figs.~\ref{fig:setup}(b) and \ref{fig:setup}(c). Adopting the terminology used in the context of triangular always-on EO qubits~\cite{weinstein_energetic_2005,Broz2026}, we refer to this point as the `leakage-protected idle' (LPI) point of the SAGE qubit.

The energy spectrum in Fig.~\ref{fig:setup}(b) shows that the point $\mu_i=0$ for all $i$ corresponds to a full charge noise sweet spot, where the qubit splitting $\epsilon_q$ is to first order insensitive to chemical potential fluctuations, $\partial\epsilon_q/\partial\mu_i=0$. This is similar to the charge noise sweet spot that was previously discussed for the three-dot always-on exchange-only (AEON) qubit~\cite{shim_charge-noise-insensitive_2016}. On the other hand, similar to other always-on EO qubits, the SAGE qubit remains linearly susceptible to fluctuations in the tunnel couplings, see Fig.~\ref{fig:setup}(c).

Throughout this paper, unless specified otherwise, we will always operate the SAGE qubit at the charge noise sweet spot, i.e., we will set all dot chemical potentials to zero. Any SAGE single-qubit rotation can then be performed by decreasing the appropriate exchange couplings by reducing the effective tunneling strengths between neighboring dots, see Fig.~\ref{fig:setup}(d).

\subsection{Noise model}
\label{subsec:noise}

The primary sources of decoherence in exchange-controlled qubits are magnetic field gradients and charge fluctuations. We model the magnetic field gradients as random quasistatic fluctuations of the local magnetic field,
\begin{equation}
H_B=\sum_{i,\sigma,\sigma'} h_i c_{i\sigma}^\dagger (s_i^z)^{\sigma\sigma'}c_{i\sigma'},
\end{equation}
where $h_i$ for $i\in\{1,2,3,4\}$ is sampled uniformly from the interval  $[-\delta h,\delta h]$ and $s_i^z$ is the Pauli $z$ matrix acting on the spin of the electron on dot $i$. Note that we take the magnetic disorder to be polarized along the $z$ direction since we are working in a large background magnetic field. Magnetic disorder polarized along the $x$ or $y$ direction would mix sectors with different total $S_z$ quantum numbers, but these processes are strongly suppressed due to the large energy splitting generated by the background field. The total Hamiltonian in the presence of magnetic noise is then $H=H_0+H_B$.

\begin{figure}[tb]
  \centering
  \includegraphics[width=0.8\linewidth]{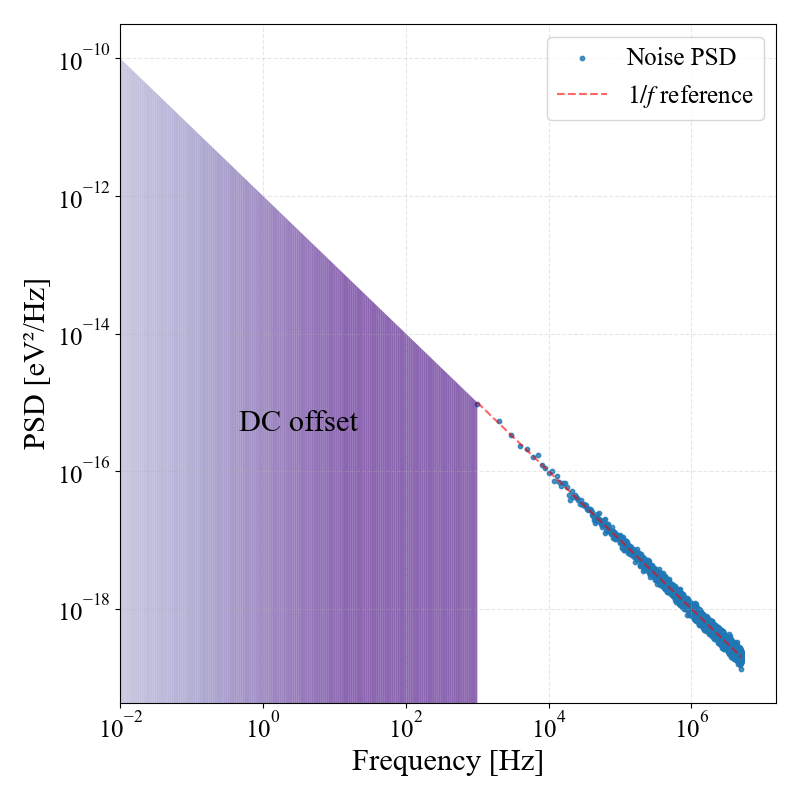}
  \caption{Energy-fluctuation PSD averaged over 100 realizations of $1/f$ noise with $A_\mu = 1 ~\mu$eV. Here, we have included a low-frequency cutoff at $1$~kHz, below which the noise is treated as DC noise.}
  \label{fig:noise}
\end{figure}

Charge noise is incorporated into our model via time-dependent fluctuations of the gate voltages. We model these fluctuations as $1/f$ noise with a power spectral density (PSD) of
\begin{equation}
S_V(f)=A_V^2/|f|,
\end{equation}
where $A_V$ determines the strength of the noise. We consider both fluctuations $\delta V_{i}(t)$ of the plunger gate voltages $V_i$ and fluctuations $\delta V_{ij}(t)$ of the exchange gate voltages $V_{ij}$. The former lead to fluctuations in the chemical potential of the individual dots, $\mu_{i}^\mathrm{noisy}(t)=\mu_{i}+\delta \mu_{i}(t)=\alpha(V_{i}+\delta V_{i}(t))$, while the latter lead to time-dependent fluctuations of the tunneling amplitudes $t_{ij}^\mathrm{noisy}(t)=t_{ij}\exp( \delta V_{ij}(t)/\beta)$. These effects are incorporated into the total Hamiltonian via $\mu_{i}\rightarrow \mu_{i}^\mathrm{noisy}(t)$ and $t_{ij}\rightarrow t_{ij}^\mathrm{noisy}(t)$ in Eq.~(\ref{eq:hubbard}). To facilitate comparison with experimentally reported noise amplitudes, we also introduce the PSD in terms of energy fluctuations $S_\mu(f)=A_\mu^2/|f|$, where $A_\mu=\alpha A_V$. In Fig.~\ref{fig:noise}, we show an example of $S_\mu(f)$ averaged over 100 realizations of numerically generated $1/f$ noise. (Here, we have included a low-frequency cutoff at $1$~kHz, below which the noise is treated as DC noise.) Unless specified otherwise, the noisy Hubbard model---incorporating both magnetic fluctuations $H_B$ as well as charge-noise induced gate voltage fluctuations as described above---will serve as the starting point for all our numerical simulations presented in the following sections.

\begin{figure*}[tb]
  \centering
  \includegraphics[width=0.8\linewidth]{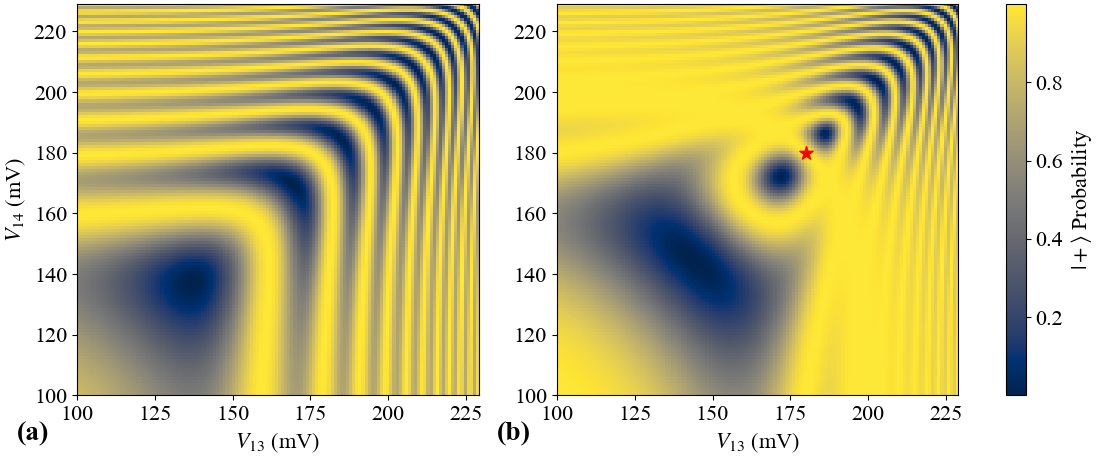}
  \caption{Return probability of measuring the $|+\rangle$ state after two simultaneous exchange pulses of length $\sim\!40$~ns as a function of exchange gate voltages. (a) Decoupled second dot, $t_{12}=0$. In this case, we recover the standard `fingerpinch' plot for the return probability in a three-dot AEON qubit. (b) SAGE qubit with $t_{12}\sim0.009$~meV ($V_{12}=180$~mV). Here, the SAGE idle point $V_{12}=V_{13}=V_{14}$ is indicated by a red star. When both exchange interactions are increased compared to the idle point (upper right corner of the plot), we find that the oscillation pattern is qualitatively the same as for the AEON qubit in (a). The three other sectors correspond to scenarios where one or both of the exchange couplings are decreased compared to the idle point. The dot chemical potentials were set to zero in both panels and we use $U=1$~meV.}
  \label{fig:fingerpinch}
\end{figure*}

\section{Single-qubit operation}
\label{sec:singlequbit}

\subsection{Exchange characterization}

Experimentally, the exchange interaction between the quantum dots forming an EO qubit is typically calibrated by measuring the return probability of an encoded state as a function of two orthogonal voltage-controlled axes (e.g., exchange gate voltages $V_{ij}$ or plunger gate voltages $V_i$). This gives rise to the standard fingerprint or fingerpinch maps frequently encountered in the experimental literature~\cite{reed_reduced_2016,ha_flexible_2022,acuna_coherent_2024,Madzik2025,Broz2025,Broz2026}. To facilitate comparison with future experiments, we present examples of such return probability maps for SAGE qubits in an ideal noiseless situation.

In Fig.~\ref{fig:fingerpinch}, we show the return probability of the encoded $|+\rangle$ state after two simultaneous exchange pulses on $V_{13}$ and $V_{14}$ of length $\sim 40$~ns. Here, we model the corresponding tunnel amplitudes with $20$~ns plateau duration and a $10$~ns linear ramp time at each end. To make a connection with the existing literature, we first consider the situation where the second dot is effectively decoupled, $t_{12}=0$. In this case, our system becomes effectively equivalent to the AEON qubit introduced in Ref.~\cite{shim_charge-noise-insensitive_2016} and experimentally studied in Ref.~\cite{Broz2025}. Explicitly, the AEON qubit Hamiltonian can be obtained from our Eq.~(\ref{eq:qubit}) by setting $J_{12}=0$. The AEON qubit is operated at equal and nonzero exchange couplings $J_{13}=J_{14}$ during idle, defining a gapped qubit with computational basis states aligned along the $z$ direction. In Fig.~\ref{fig:fingerpinch}(a), we reproduce the standard `fingerpinch' plot for the return probability that is expected for this type of qubit~\cite{Broz2025}. Here, $P_{|+\rangle}$ oscillates as a function of the exchange pulse strength, with a translation along the diagonal $V_{13}=V_{14}$ corresponding to a rotation around the $z$ axis.

Next, in Fig.~\ref{fig:fingerpinch}(b), we show results for a SAGE qubit with $t_{12}\sim0.009$~meV (corresponding to $V_{12}=180$~mV). Importantly, the additional tunnel coupling renders the spectrum of the SAGE qubit gapless at $t_{12}=t_{13}=t_{14}\equiv t_0$. This is the SAGE qubit's idle point [indicated by the red star in Fig.~\ref{fig:fingerpinch}(b)], where the qubit can safely idle in any initial state despite the exchange couplings being always-on. As we tune the tunnel couplings away from the idle point, we again find that the return probability $P_{|+\rangle}$ oscillates as a function of the exchange pulse strength, with the diagonal corresponding to a rotation around the $z$ axis as expected from Eq.~(\ref{eq:qubit}). However, in contrast to the AEON qubit (which does not have a gapless point unless the couplings are turned off altogether), it is now possible to tune the exchange interactions away from the idle point in two directions. The upper right corner of Fig.~\ref{fig:fingerpinch}(b), where both exchange interactions are increased compared to the idle point, qualitatively resembles the AEON fingerpinch plot shown in Fig.~\ref{fig:fingerpinch}(a). The three other sectors correspond to the cases where one or both of the exchange couplings are decreased compared to the idle point. The SAGE qubit can therefore idle at an operating point of maximum leakage suppression, while single-qubit gates are performed by asymmetrically lowering the exchange couplings to nonzero values, thereby protecting the energy gap between leakage and computational states.

\subsection{Coherence times and dynamical decoupling strategies}

Even when operated at the charge noise sweet spot discussed in Sec.~\ref{subsec:hubbard}, SAGE qubits suffer from an increased sensitivity to charge noise during idle compared to conventional EO qubits. At the same time, the always-on operation leads to superior protection against magnetic noise, with SAGE qubits dramatically outperforming conventional EO qubits in systems where the noise is predominantly magnetic~\cite{Foulk2025}. As such, charge noise is the main impediment limiting SAGE qubit idle coherence times and gate fidelities. This motivates us to study the effect of charge noise, and potential mitigation strategies via dynamical decoupling, for SAGE qubits in more detail.

\begin{figure*}[tb]
  \centering
  \includegraphics[width=0.9\linewidth]{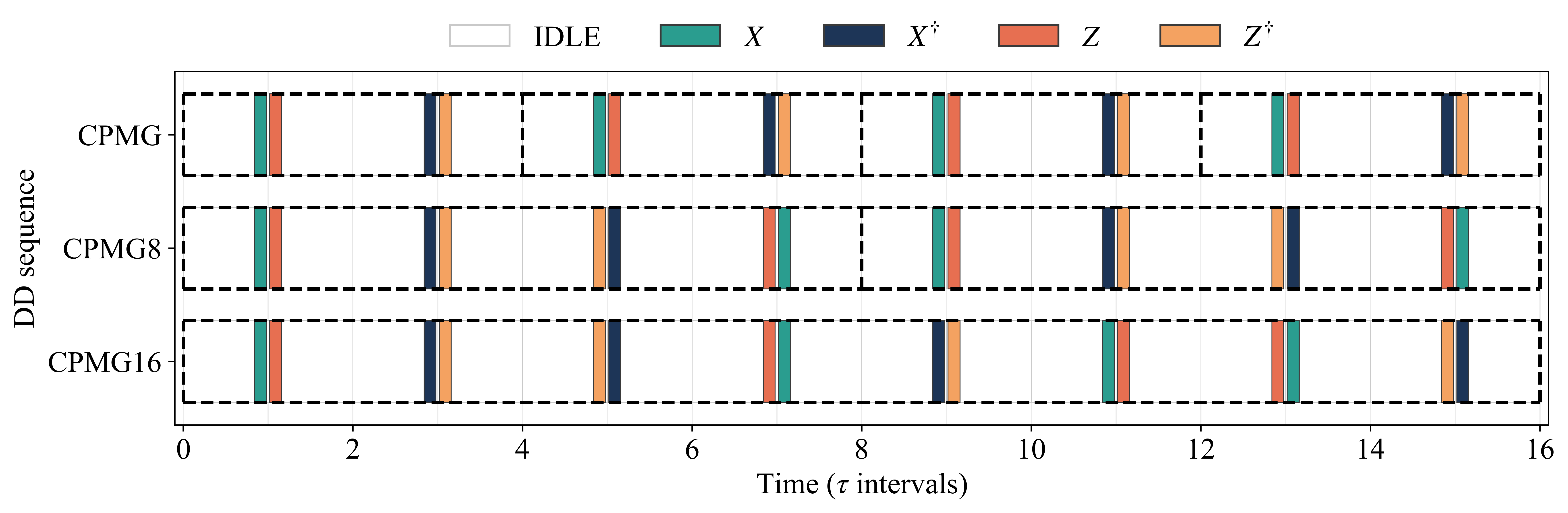}
  \caption{Dynamical decoupling pulse sequences used to mitigate the effects of charge noise during SAGE idling. Here, $X=\exp(-i\frac{\pi}{2}\sigma_x)$ and $Z=\exp(-i\frac{\pi}{2}\sigma_z)$ denote $\pi$-rotations around the qubit's $x$ and $z$ axes, respectively.}
  \label{fig:xy4}
\end{figure*}

In the following, we focus on using repeated $\pi$-pulses about the qubit $y$ axis in the spirit of CP/CPMG~\cite{Carr1954Effects,Meiboom1958Modified} to mitigate the effects of charge noise during SAGE idling. Indeed, for SAGE qubits, charge noise leads to time-dependent fluctuations in the exchange couplings, which in turn introduce $\sigma_x$ and $\sigma_z$ errors in the qubit Hamiltonian, see Fig.~\ref{fig:setup}(d). Therefore, sequences of equally spaced dynamical decoupling pulses about the $y$ axis are a suitable choice that render the time-averaged noise Hamiltonian zero to first order for sufficiently slow fluctuations. In the SAGE architecture, rotations about the $y$ axis are realized by concatenating rotations about the $x$ and $z$ axes [see again Fig.~\ref{fig:setup}(d)], which we use as the building blocks of our CPMG-like pulse sequences. Our starting point is a basic CPMG-like sequence in which the signs of the constituent $x$ and $z$ rotations are reversed in every other pulse pair, see the top row of Fig.~\ref{fig:xy4}. Furthermore, we also consider two more advanced CPMG-type sequences obtained by phase shifting and concatenating this basic CPMG sequence. Explicitly, the CPMG-8 sequence (second row of Fig.~\ref{fig:xy4}) is an eight-pulse sequence obtained by combining our basic CPMG sequence with its time-reversed copy, while the CPMG-16 sequence (third row of Fig.~\ref{fig:xy4}) consists of 16 pulses and is obtained  by combining the CPMG-8 sequence with its phase-shifted copy.

\begin{figure*}[tb]
  \centering
  \includegraphics[width=0.96\linewidth]{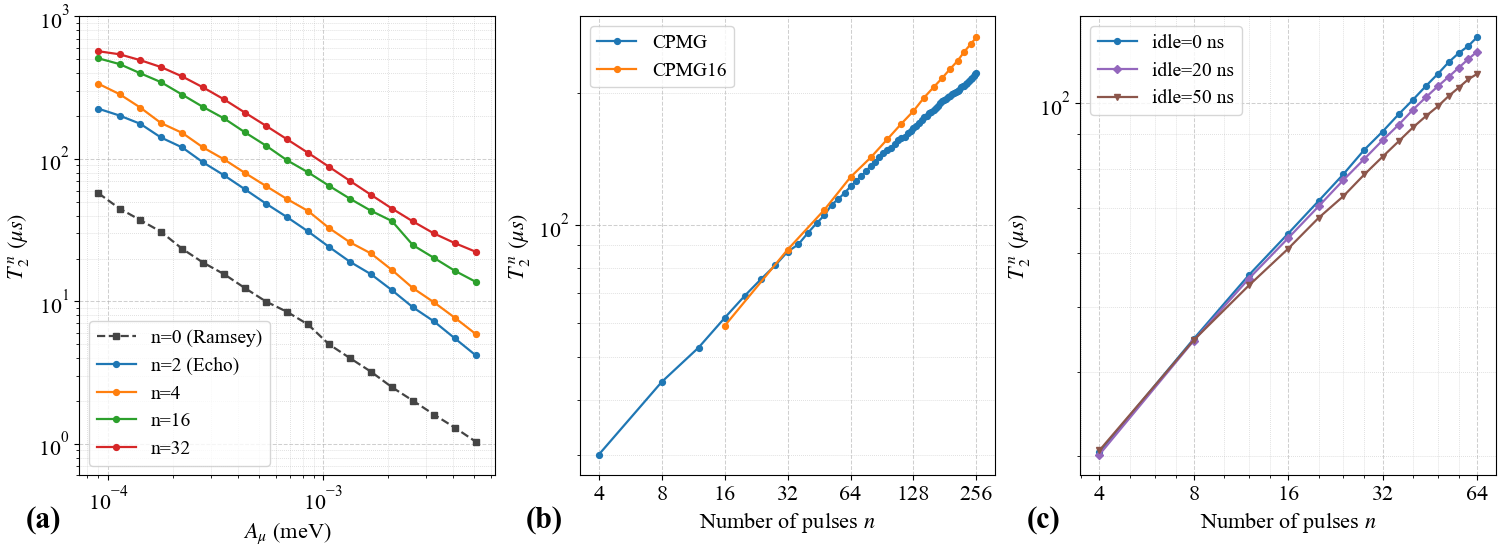}
  \caption{Effect of dynamical decoupling on SAGE idle coherence times. (a) $T^n_2$ times obtained by applying the CPMG pulse sequence shown in the top row of Fig.~\ref{fig:xy4} for different numbers of pulses $n$ as a function of charge noise strength. (b) $T_2^n$ times obtained by applying CPMG (first row of Fig.~\ref{fig:xy4}) and CPMG-16 (third row of Fig.~\ref{fig:xy4}) pulse sequences, respectively, as a function of number of pulses $n$. (c) $T^n_2$ times obtained by applying CPMG pulse sequences for varying idle times between neighboring $X$ and $Z$ pulses as a function of number of pulses $n$. In all panels, we set $t_0\sim 0.007U$ and $U=1$~meV, which corresponds to an experimentally realistic value of $J_0/h\sim50$~MHz. In (a) and (b), the idling time between neighboring $X$ and $Z$ pulses is set to zero. In (b) and (c), we set $A_\mu=1~\mu$eV, and in all panels we have $\delta h=0.2$~neV. Each data point was averaged over $400$ disorder realizations.}
  \label{fig:t2}
\end{figure*}

In Fig.~\ref{fig:t2}, we show numerically calculated SAGE $T_2^n$ coherence times obtained by applying the CPMG-type pulse sequences shown in Fig.~\ref{fig:xy4} to a qubit initially prepared in the $|+\rangle$ state. Here, $n$ denotes the total number of \textit{physical} pulses that are applied to the system, i.e., the number of equally spaced effective $\pi$-pulses about the $y$ axis is $n/2$. The coherence times are calculated by averaging the time-dependent off-diagonal element of the qubit density matrix over $400$ disorder realizations and fitting a Gaussian decay function $A \exp(-(t/T_2)^2)+B$. Throughout our simulations, we set $t_0\sim 0.007U$ and $U=1$~meV, which results in an experimentally realistic value of $J_0/h\sim50$~MHz. $\pi$-rotations about the qubit's $x$ and $z$ axes are then performed by reducing the appropriate exchange couplings so the resulting Pauli operator has energy splitting $J_0/2$. For this choice of parameters, an individual square $\pi$-pulse in our CPMG sequence has a duration of $\sim 20$~ns. Charge noise affects the tunnel couplings and dot chemical potentials both during idling as well as while pulses are being performed, such that all pulses in our simulation are noisy.

In Fig.~\ref{fig:t2}(a), we show SAGE $T^n_2$ times obtained by applying the basic CPMG pulse sequence shown in the top row of Fig.~\ref{fig:xy4} for different numbers of pulses $n$ as a function of charge noise strength. Here, the range of charge noise strengths we consider is motivated by recent experiments on silicon-based quantum dot devices, which typically report amplitude spectral densities $\sim\!1 \;\mu\textrm{eV}/\sqrt{\text{Hz}}$ at 1 Hz~\cite{Freeman2016,Connors2022}, with potential further improvements as fabrication moves to industrial foundries~\cite{Steinacker2025,elsayed_low_2022,neyens_probing_2024,Koch2025,Thomas2025}. We find that already a single echo pulse leads to a significant improvement of the SAGE idle coherence time $T^{n=2}_2$ (blue line) compared to the bare $T^{n=0}_2$ Ramsey coherence time (black line). Increasing the pulse number $n$ leads to a gradual increase in $T_2^n$ times as expected, with $T_2^{n=32}$ (red line) exceeding $T^{n=0}_2$ by over an order of magnitude across the entire range of charge noise amplitudes we consider. Explicitly, we find that $T_2^{n=32}$ is on the order of $\sim 100~\mu$s for realistic charge noise amplitudes. 

At the same time, as $n$ increases further, pulse imperfections begin to have a noticeable effect. This is illustrated in Fig.~\ref{fig:t2}(b), where we compare the $T^n_2$ times obtained from our basic CPMG sequence (first row of Fig.~\ref{fig:xy4}) and the symmetrized CPMG-16 sequence (third row of Fig.~\ref{fig:xy4}) as a function of the number of pulses $n$. For $n\gtrsim 64$, we start to see a clear deviation between the two pulse sequences resulting from the accumulation of pulse errors in the standard CPMG sequence, whereas the CPMG-16 sequence does a better job at canceling out these errors.

In the above, we have set the idle time between neighboring $X$ and $Z$ pulses (i.e., pairs of green and orange, or navy and gold pulses in Fig.~\ref{fig:xy4}) to zero. However, this may be experimentally challenging to implement. Therefore, in Fig.~\ref{fig:t2}(c), we study the effect of small idling periods between neighboring $X$ and $Z$ pulses as a function of pulse number $n$. We find that short interpulse idle times only weakly affect the performance of our CPMG pulse sequence, leading to only a slight decrease in the observed coherence times that becomes more pronounced as $n$ increases. As such, the CPMG pulse sequences proposed here remain applicable even if it is not possible to perform extremely closely spaced $X$ and $Z$ pulses.

We note that since the SAGE qubit is a gapless qubit, there is no natural distinction between $T_1$ and $T_2$ times. Here, we have studied the dephasing time of a state initially prepared in the qubit's $x$ basis as a simple measure of qubit coherence. However, it is worth mentioning that different initial states can generally exhibit slightly different dephasing times since both our noise model and our dynamical decoupling sequence are asymmetric in the $xy$ plane.

\section{Two-qubit gates}
\label{sec:twoqubit}

\begin{figure}[tb]
  \centering
  \includegraphics[width=0.9\columnwidth]{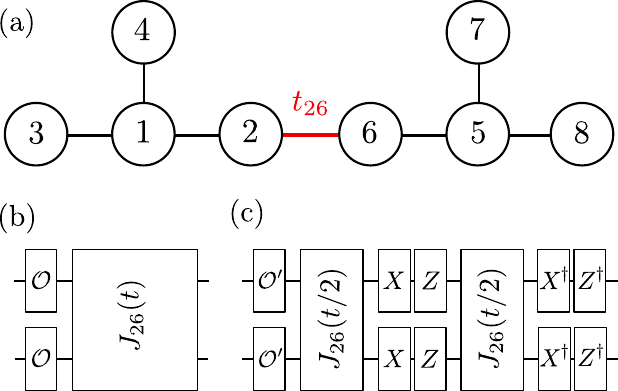}
  \caption{(a) Geometry for SAGE two-qubit gates. A single interqubit exchange pulse is performed by turning on the tunnel coupling $t_{26}$ between dots $2$ and $6$ (red line) while keeping all intraqubit tunnel couplings at a fixed value $t_0\gg t_{26}$ (black lines). (b) Schematic illustration of the pulse sequence to perform a SAGE CZ gate as introduced in Ref.~\cite{Foulk2025}. Here, $\mathcal{O}$ denotes a single-qubit rotation $\mathcal{O}=\exp(\frac{i\pi \theta}{2}\sigma_z)$ with $\theta \approx 3/8+9J_c/128J_0$. (c) Schematic illustration of the pulse sequence to perform a SAGE CZ gate with an echo pulse inserted halfway through the interqubit pulse. Here, $\mathcal{O}'=\exp(\frac{i\pi}{4}\sigma_z)$, and $X$ and $Z$ were defined in Fig.~\ref{fig:xy4}. For simplicity, the schematics in (b) and (c) assume square pulses; in our simulations, finite ramp times are included and the pulses are calibrated to produce the desired effective pulse areas.}
  \label{fig:twoqubit}
\end{figure}

SAGE two-qubit gates can be performed by turning on a single interqubit exchange pulse between two neighboring qubits as shown in Fig.~\ref{fig:twoqubit}. Specifically, a fully entangling gate can be realized by turning on the interqubit coupling $t_{26}$ for a fixed amount of time while setting all intraqubit couplings to a fixed value $t_{12}=t_{13}=t_{14}=t_{56}=t_{57}=t_{58}\equiv t_0\gg t_{26}$~\cite{Foulk2025}. Using a Schrieffer-Wolff transformation, one can show that pulsing $J_{26}\equiv J_c$ results in an effective two-qubit interaction
\begin{equation}
H_\textrm{2q} = J_\mathrm{eff}\sigma^z_1\sigma^z_2 - J_s (\sigma_1^z + \sigma_2^z),
\label{eq:twoqubit}
\end{equation}
where $J_\mathrm{eff}\approx J_c^2/6J_0 + 5J_c^3/32J_0^2$ and $J_s\approx J_c^2/24J_0 + J_c^3/64J_0^2$ up to third order in $J_c/J_0$. Thus, pulsing $J_c$ such that $\int dt J_\mathrm{eff}(t)/\hbar=\pi/4$ will realize a CZ gate up to local unitaries, which can be removed by performing appropriate single-qubit gates, see Fig.~\ref{fig:twoqubit}(b) for a schematic illustration.

\begin{figure*}[tb]
  \centering
  \includegraphics[width=0.32\linewidth]{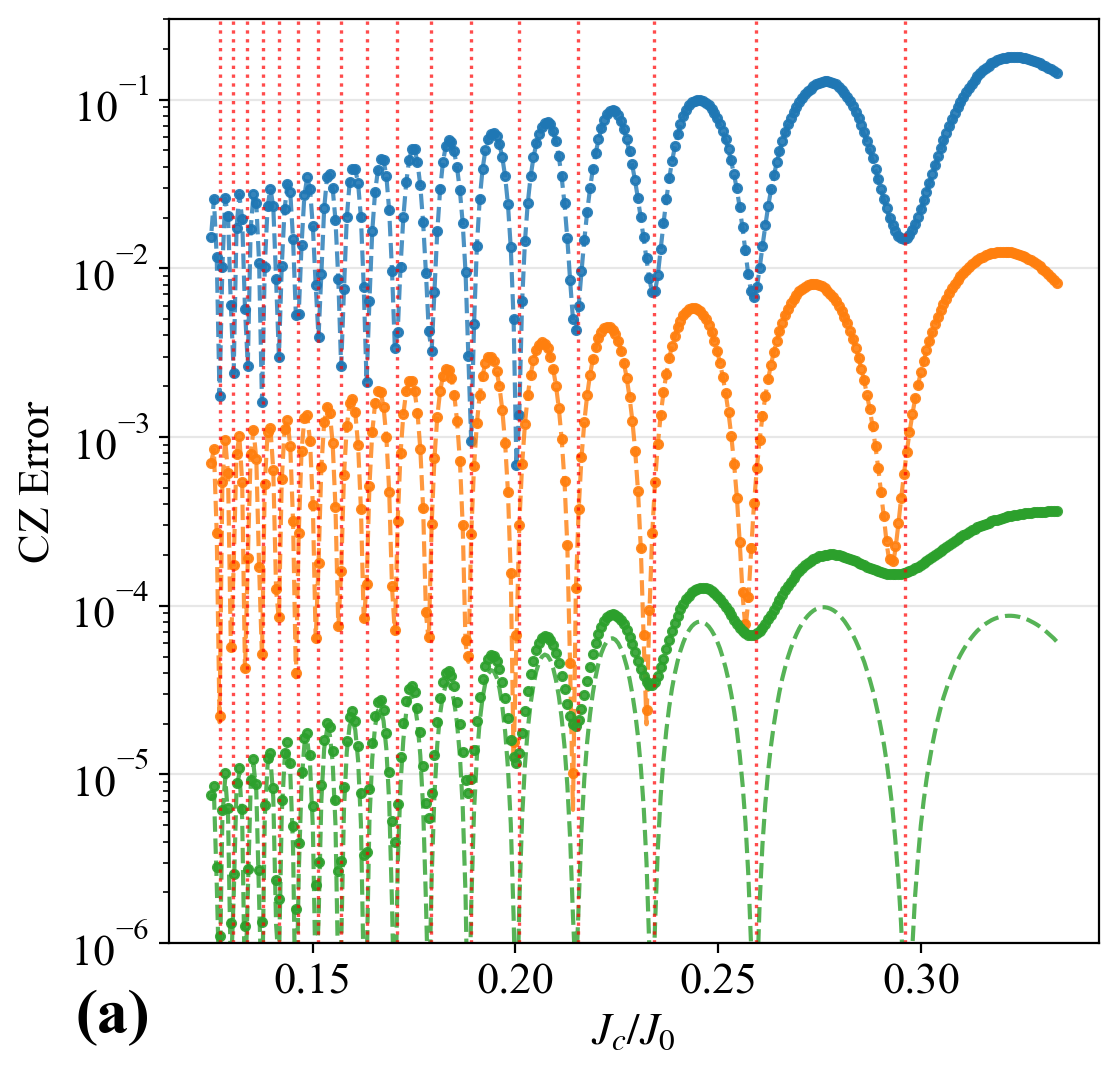}
  \includegraphics[width=0.32\linewidth]{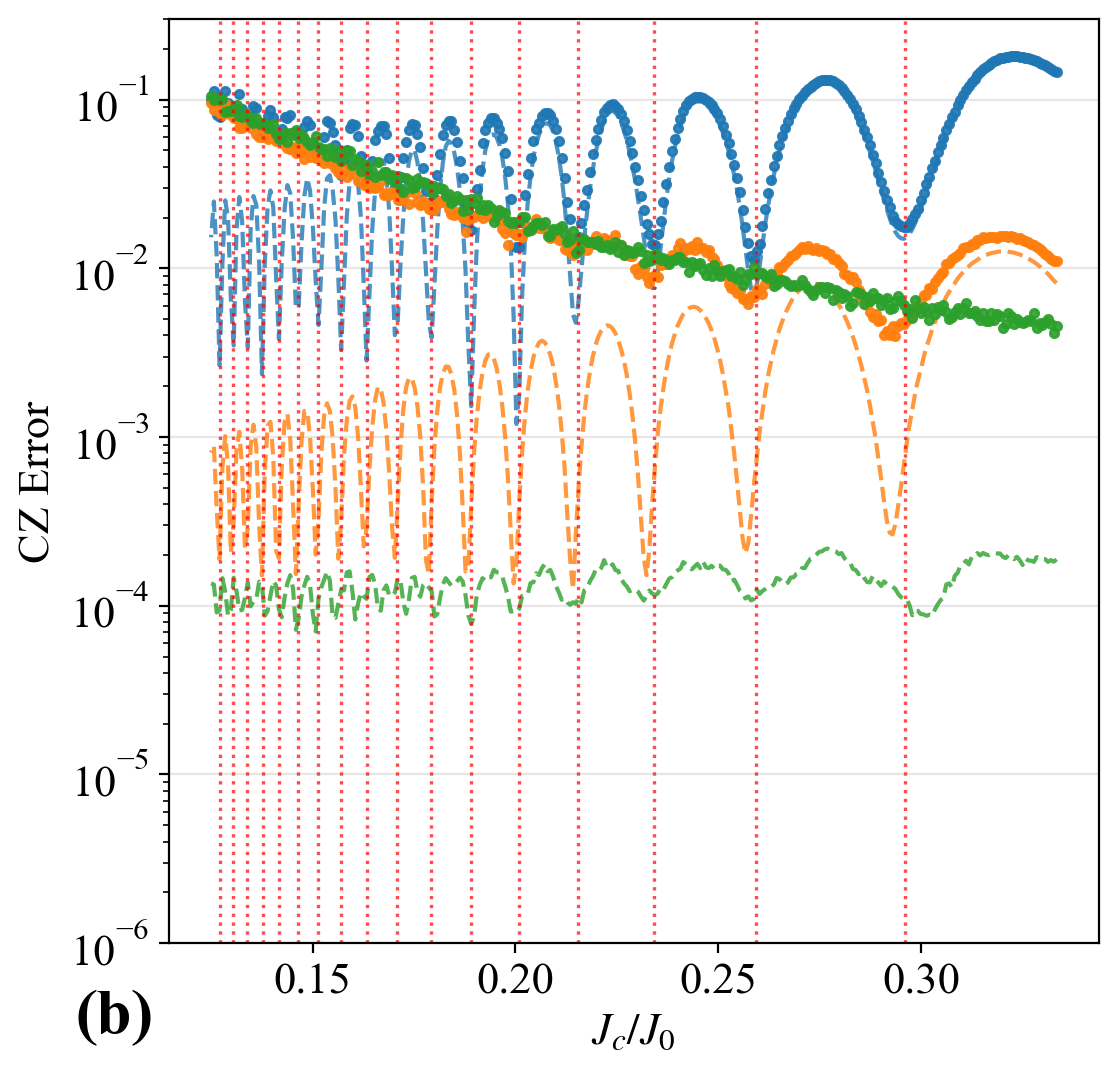}
  \includegraphics[width=0.32\linewidth]{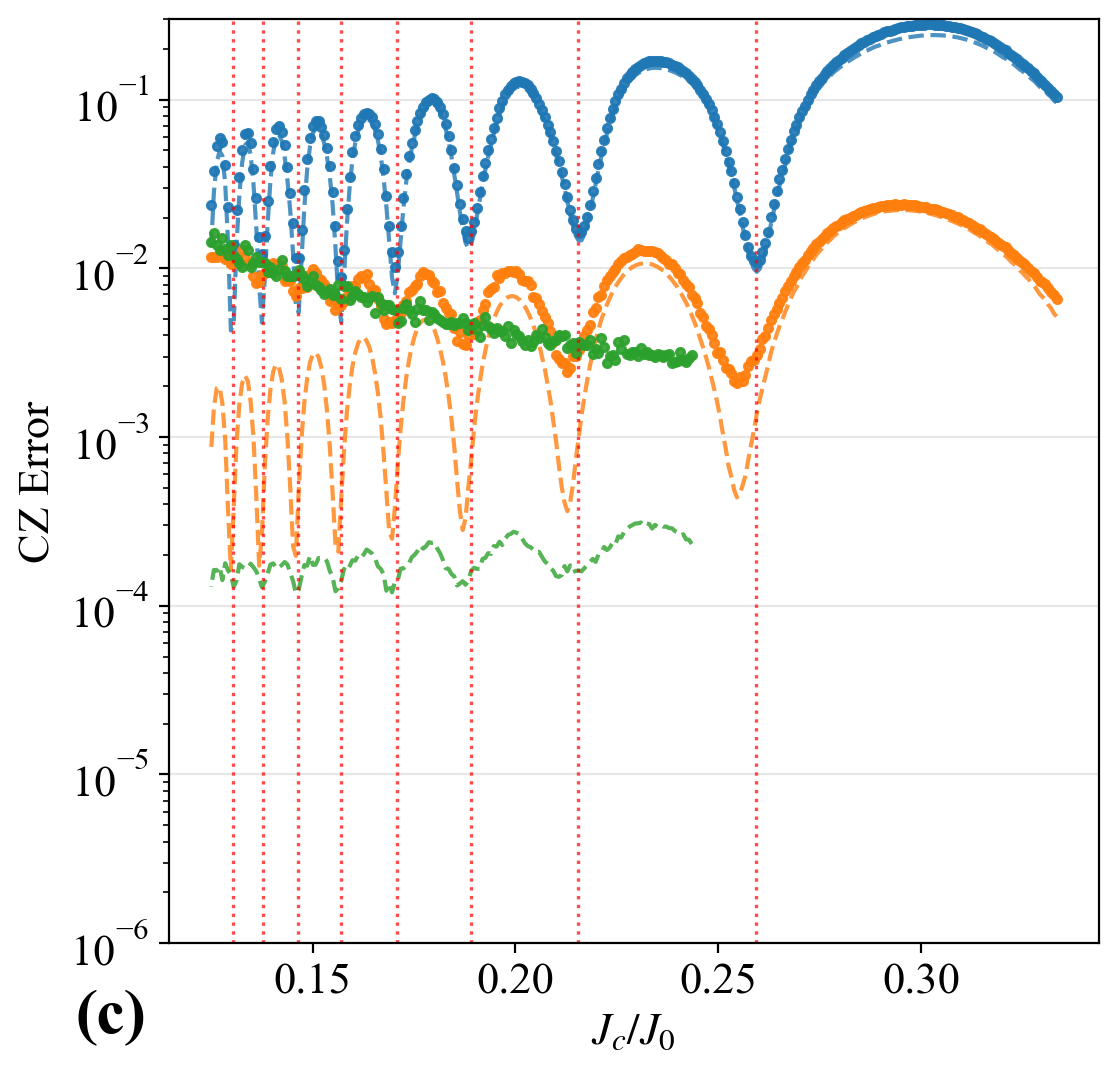}
  \caption{Infidelity (dots) and leakage (dashed lines) of SAGE CZ gate as a function of $J_{c}/J_0$ for different ramp times of the interqubit pulse (green: $100$~ns, orange: $50$~ns, blue: $10$~ns). (a) Noiseless system. Here, small values of $J_{c}/J_0$ and long pulse ramp times yield the highest fidelities due to the large energy gap between computational and leakage states and adiabatic suppression of leakage. (b) Noisy system with charge noise and magnetic noise of strength $A_\mu=1~\mu$eV and $\delta h=0.2$~neV, respectively, without echo pulse. In this case, small ratios $J_{c}/J_0$ are noise-limited because the corresponding gate times are long, while larger ratios enter a crossover to an intrinsic-leakage-limited regime in which sufficiently long ramp times suppress leakage oscillations and improve the fidelity. (c) Same noisy system as in panel (b), but with a single echo pulse inserted after the interqubit pulse has accumulated half the target entangling angle. The echo pulse suppresses low-frequency computational errors and improves the fidelity in large parts of the parameter space considered here. Simulations are based on a Heisenberg model and each data point is averaged over $100$ disorder realizations. The leakage oscillations can be  understood analytically (see Appendix~\ref{app:oscillations}), and their expected minima in the absence of ramping are indicated by  vertical red dotted lines.
}
  \label{fig:ramptime}
\end{figure*}

\begin{figure*}[tb]
  \centering
  \includegraphics[width=0.85\linewidth]{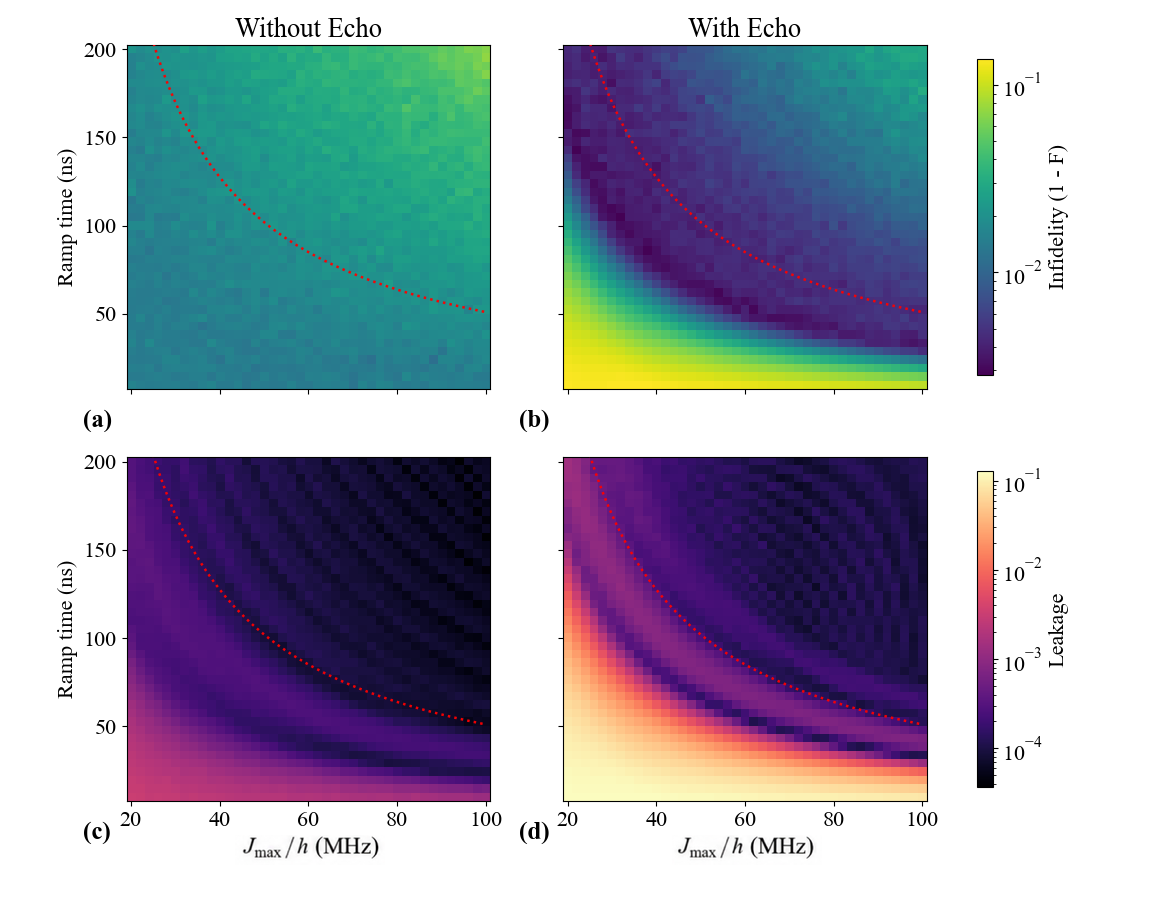}
    \caption{(a,b) Infidelity and (c,d) leakage for SAGE CZ gate (a,c) without and (b,d) with echo pulse as a function of pulse ramp time $t_\mathrm{ramp}$ and maximally achievable exchange coupling $J_\mathrm{max}$. The ratio $J_{c}/J_0=J_{c}/J_\mathrm{max}=1/5$ is kept fixed. We set $\delta h=0.2$~neV and $A_\mu=1~\mu$eV in all panels. Simulations are based on a Heisenberg model and each data point is averaged over 100 disorder realizations. The dashed red lines in this parameter space correspond to a ramp time $t_\textrm{ramp} = 80 \;\tau_0$ to ensure adiabatic operation (see Appendix~\ref{app:adiabatic}).}
  \label{fig:pulseramp_vs_maxJ}
\end{figure*}

A key difficulty in implementing two-qubit gates for exchange-only qubits is that the exchange interaction between qubits does not preserve the computational subspace, such that leakage can occur intrinsically even in the absence of noise. This necessitates either carefully designed pulse sequences to cancel leakage (e.g., Fong–Wandzura–type constructions for conventional EO qubits~\cite{fong_universal_2011,weinstein_universal_2023,heinz_fast_2024}) or energetic suppression mechanisms such as in resonant exchange (RX)~\cite{doherty_two-qubit_2013}, AEON~\cite{shim_charge-noise-insensitive_2016}, and SAGE qubits~\cite{Foulk2025}. Explicitly, for SAGE two-qubit gates, the always-on intraqubit couplings ensure that leakage states are separated from the computational subspace by an energy gap $\propto J_0$. A sufficiently slow and smooth ramp of the interqubit pulse relative to the leakage gaps then ensures that the state evolves adiabatically, suppressing the population of leakage states. In the following, we model the interqubit pulse as a square pulse with raised-cosine edges constructed from a reference square pulse with amplitude $J_c$. Specifically, starting from the square pulse, we introduce raised-cosine rising and falling edges of duration $t_{\mathrm{ramp}}$, reduce the flat plateau duration by one ramp time, and then readjust the plateau height such that the desired effective pulse area is maintained. The fidelity of SAGE two-qubit gates generally depends on all three parameters $J_c$, $J_0$, and $t_\mathrm{ramp}$, as well as on their interplay.

In order to explore this rather large parameter space in an efficient way, we first present results obtained directly from the Heisenberg Hamiltonian [see Eq.~(\ref{eq:heisenberg})] before moving to simulations based on the Hubbard model [see Eq.~(\ref{eq:hubbard})] further below. (We have checked that both approaches yield consistent results.) We begin by analyzing the effects of pulse ramping by calculating the infidelity\footnote{We consider the Haar-averaged fidelity $\mathcal{F}= [|\mathrm{Tr}(U^\dagger V)|^2 + \mathrm{Tr}(V^\dagger V)] / [d (d + 1)]$, where $d=4$ for two-qubit gates, $U$ is the target unitary, and $V$ is the actually simulated map projected onto the logical subspace (which is generally non-unitary in the presence of leakage)~\cite{Pedersen2007}. The corresponding infidelity is $1-\mathcal{F}$.} (dots) and leakage\footnote{We define the leakage rate $p_\mathrm{leak}= 1- \mathrm{Tr}(V^\dagger V) / d  $, where $d=4$ for two-qubit gates and $V$ is the simulated map projected onto the logical subspace.} (dashed lines) for the SAGE CZ gate as a function of $J_c/J_0$ for three representative pulse ramp times $t_\mathrm{ramp}=10$~ns, $50$~ns, and $100$~ns, see Fig.~\ref{fig:ramptime}. The always-on intraqubit exchange is kept fixed to $J_0/h=50$~MHz.

Figure~\ref{fig:ramptime}(a) shows the results for an ideal noiseless system. For short ramp times, we observe relatively high infidelities that oscillate strongly as a function of $J_{c}/J_0$. This infidelity is due to `intrinsic' leakage coming from non-adiabatic population of leakage states. On the other hand, for long pulse ramp times, the achievable fidelities are several orders of magnitude higher and leakage oscillations are suppressed. In this regime, computational errors due to the truncation of the perturbative expansion underlying Eq.~(\ref{eq:twoqubit}) become the relevant source of errors as $J_c/J_0$ increases (see the green line).

For realistic noisy systems, the situation becomes more complicated as multiple competing effects are at play. This is illustrated in Fig.~\ref{fig:ramptime}(b), which is the same as Fig.~\ref{fig:ramptime}(a) but includes charge and local magnetic noise of strength $A_\mu=1~\mu$eV and $\delta h=0.2$~neV, respectively. Here, we find that small ratios $J_{c}/J_0$ result in low fidelities regardless of ramp time since the corresponding gate times are long, leading to an accumulation of noise-induced errors over time. As such, the fidelity is `noise-limited' in this regime. As $J_{c}/J_0$ increases, the effects of pulse ramping become increasingly important, leading to a crossover between a `noise-limited' and an `intrinsic-leakage-limited' regime as the gate becomes faster. This can be seen from the blue and orange curves in Fig.~\ref{fig:ramptime}(b), which begin to gradually approach their strongly oscillating noiseless counterparts as $J_{c}/J_0$ becomes sufficiently large. For the green curve in Fig.~\ref{fig:ramptime}(b), on the other hand, intrinsic leakage is so strongly suppressed that we remain in the noise-limited regime for the entire range of $J_{c}/J_0$ considered here.

Next, we consider a simple refocusing scheme that can be used to mitigate the effects of single-qubit charge noise on the fidelity of SAGE two-qubit gates. Specifically, we apply an effective $\pi$ echo pulse about the $y$ axis (i.e., two consecutive $\pi$ pulses about the $x$ and $z$ axes) to both qubits after the interqubit pulse has accumulated half the target entangling angle; see Fig.~\ref{fig:twoqubit}(c) for a schematic illustration in the case of square pulses.
The effect of this echo pulse is twofold: First, it echoes out the single-qubit terms $\propto J_s$ in Eq.~(\ref{eq:twoqubit}), leaving us with a pure $\sigma_z^{(1)}\sigma_z^{(2)}$ interaction. Therefore, the single-qubit rotations required to perform a CZ gate reduce to $\pi/2$ rotations around the $z$ axis, see again Fig.~\ref{fig:twoqubit}(c). Second, the echo pulse cancels out low-frequency charge noise similar to the single-qubit CPMG pulses discussed in Sec.~\ref{sec:singlequbit}. This effect can be seen in Fig.~\ref{fig:ramptime}(c), where the echo pulse improves the fidelity of the SAGE CZ by about an order of magnitude at small $J_{c}/J_0$. In this regime, regardless of pulse ramp time, the main source of errors are computational errors due to $1/f$ charge noise, the low-frequency part of which can be efficiently echoed out by a refocusing pulse. For intermediate to short pulse-ramp times (orange and blue curves), we observe that the crossover from the noise-limited to the intrinsic-leakage-limited regime is now shifted to smaller values of $J_{c}/J_0$ compared to Fig.~\ref{fig:ramptime}(b) since the infidelity due to computational errors is reduced by the echo pulse. We note that for the longest pulse ramp time considered here ($t_\mathrm{ramp}=100$~ns, green line) we only consider interqubit exchange couplings up to a critical value $J_c/J_0\sim 0.25$, since beyond that point the interqubit pulse cannot be fully ramped up without the pulse area $\int dt J_\mathrm{eff}(t)/\hbar$ becoming larger than its target value of $\pi/8$.

To identify optimal operating regimes for SAGE two-qubit gates for a given ratio $J_c/J_0$, we additionally show color maps for the infidelity and leakage of the SAGE CZ as a function of pulse ramp time and maximally achievable exchange coupling $J_\mathrm{max}$ in Fig.~\ref{fig:pulseramp_vs_maxJ}. Here, $J_\mathrm{max}$ sets the strength of the intraqubit exchange coupling, $J_\mathrm{max}=J_0$, which we have kept fixed so far (see Fig.~\ref{fig:ramptime}). Now, we instead keep the ratio $J_c/J_0=1/5$ fixed. First, in Figs.~\ref{fig:pulseramp_vs_maxJ}(a,c), we show the infidelity and leakage in the absence of an echo pulse. We find that the infidelity map [Fig.~\ref{fig:pulseramp_vs_maxJ}(a)] exhibits very little structure, with infidelities remaining $>1\%$ throughout the considered parameter range. This reflects the fact that we are in the noise-dominated regime, where the fidelity is limited by computational errors. Indeed, the leakage map [Fig.~\ref{fig:pulseramp_vs_maxJ}(c)] shows that leakage is negligibly small compared to the overall infidelity. This is mainly a consequence of the fact that our choice of $J_c/J_\mathrm{max}$ corresponds to a local minimum in the leakage oscillations seen in Fig.~\ref{fig:ramptime}(a,b) and therefore constitutes an optimal operating point when it comes to leakage suppression. Furthermore, we find that the leakage map has an oscillating structure in $(t_\mathrm{ramp},J_\mathrm{max})$ space, with leakage minima appearing along lines where $t_\mathrm{ramp}\propto 1/J_\mathrm{max}$ (see the red dashed line for an example). Overall, as expected, leakage decreases as we move towards longer pulse ramp times and larger $J_\mathrm{max}$.

In Figs.~\ref{fig:pulseramp_vs_maxJ}(b,d), we show the infidelity and leakage of the SAGE CZ gate in the presence of an echo pulse [see Fig.~\ref{fig:twoqubit}(c)]. For short ramp times, the total infidelity is now generally of the same order of magnitude as the leakage, indicating that leakage errors are now a dominant source of infidelity. This is mainly due to the fact that our choice of $J_c/J_0=1/5$ does not correspond to an optimal leakage suppression point (i.e., a local minimum of the leakage oscillations) in the presence of an echo pulse, but in fact to a local maximum, as shown in Fig.~\ref{fig:ramptime}(c). As such, leakage in Fig.~\ref{fig:pulseramp_vs_maxJ}(d) is generally higher than in the case without an echo pulse shown in Fig.~\ref{fig:pulseramp_vs_maxJ}(c). Nevertheless, there exist paths of ramp times $t_\mathrm{ramp} \propto 1/J_\mathrm{max}$ along which leakage remains strongly suppressed through adiabatic operation (see again the dashed red line for an example). In the adiabatic regime, the infidelity shown in Fig.~\ref{fig:pulseramp_vs_maxJ}(b) is significantly reduced compared to the case without echo pulse shown in Fig.~\ref{fig:pulseramp_vs_maxJ}(a), as the echo pulse cancels out computational errors caused by low-frequency charge noise. Pulse ramping therefore becomes even more important in the presence of an echo pulse. 

\begin{figure*}[tb]
  \centering
  \includegraphics[width=0.75\linewidth]{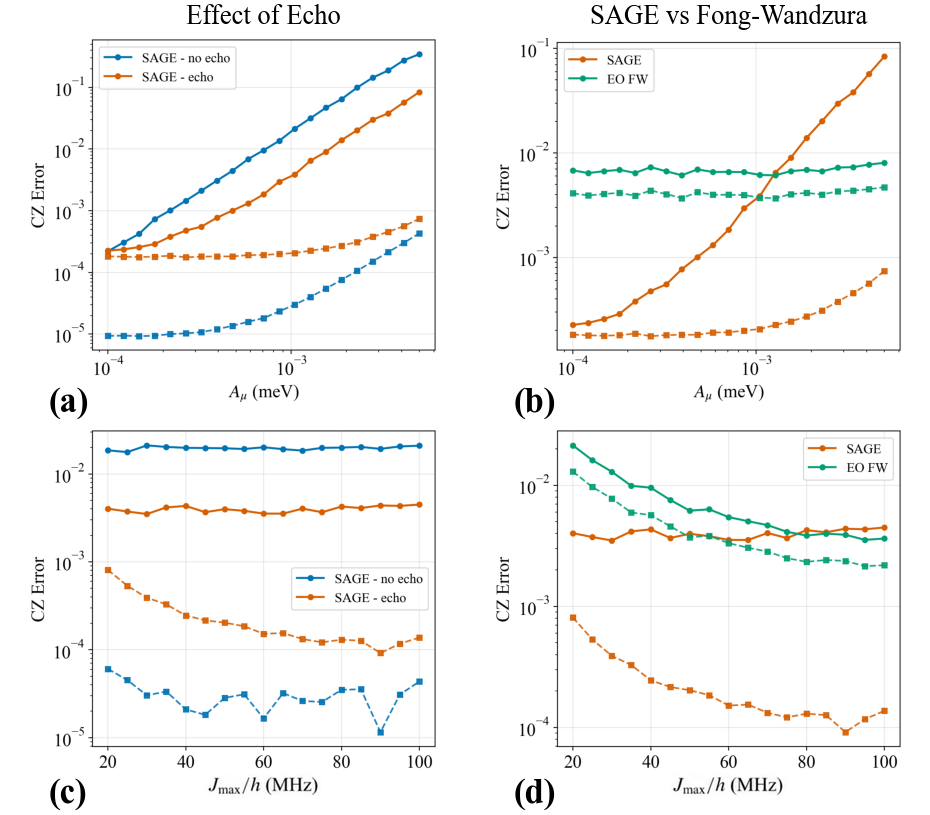}
    \caption{(a) SAGE two-qubit CZ infidelity and leakage as a function of $1/f$ charge noise strength $A_\mu$ with (red) and without (blue) echo pulse. In the adiabatic regime, SAGE two-qubit fidelities are dominated by single-qubit errors due to charge noise, such that fidelities are improved by local echo pulses. (b) SAGE CZ (with an echo pulse) compared to the Fong-Wandzura CZ for conventional EO qubits. Charge noise amplitudes bear little impact on Fong-Wandzura fidelities, which are limited by magnetic-gradient-induced leakage. 
    (c) SAGE two-qubit CZ infidelity and leakage with (red) and without (blue) echo pulse as a function of the maximal achievable exchange coupling $J_\mathrm{max}$, which sets the always-on intraqubit exchange coupling $J_0$. (d) SAGE two-qubit CZ compared to the Fong-Wandzura CZ for conventional EO qubits as a function of $J_\mathrm{max}$, which sets the pulse strength of the individual pulses in the Fong-Wandzura pulse sequence. In all panels, infidelities (leakage) are given with solid (dashed) lines. We set $\delta h=0.2$~neV in all panels, and in (c,d) the strength of charge noise is $A_\mu=1~\mu$eV. For the SAGE qubit, the ratio $J_{c}/J_\mathrm{max}=J_{c}/J_0$ is kept fixed. In the FWCZ sequence for the conventional EO qubit, consecutive pulses are separated by a $5$~ns buffer. Each data point is averaged over 200 disorder realizations, using a six-dot Hubbard model for the conventional EO qubit and an eight-dot Hubbard model for SAGE, both with $U = 3$~meV.
    }
  \label{fig:2q_vs_sigmaV}
\end{figure*}

We also note that the ratio $J_{c}/J_0$ could be optimized so that leakage is minimized both with and without echo pulse. Nevertheless, even for the ``worst-case'' scenario of suboptimal tuning considered here ($J_c/J_0 = 1/5$), extending the pulse ramp duration shifts the system further into adiabatic operation and reduces the importance of the precise choice of exchange ratio. The upshot is that, as long as the interqubit pulse is ramped up adiabatically, a single echo pulse already leads to an overall improvement of the gate fidelity even without any additional leakage optimization, which may be challenging to realize experimentally.

We also study the SAGE CZ fidelity as a function of $1/f$ charge noise strength, which we have kept fixed up to now. For this part, we verify our results using the full Hubbard model rather than the effective Heisenberg model, which allows us to incorporate charge noise in the microscopic parameters $\mu_i$ and $t_{ij}$ (both controlled via the corresponding gate voltages) rather than just in the effective exchange interactions. In Fig.~\ref{fig:2q_vs_sigmaV}(a), we show the infidelity of the SAGE CZ gate as a function of $1/f$ charge noise strength both with and without an echo pulse. Here, we work deep in the adiabatic regime at a fixed pulse ramp time $t_\mathrm{ramp}=100$~ns to ensure intrinsic leakage is suppressed. In this regime, the infidelity is dominated by single-qubit computational errors due to charge noise, the low-frequency components of which can be effectively echoed out by our echo pulse. Indeed, we find that the echo pulse reduces the total infidelity by almost an order of magnitude, completely canceling out the effects of computational errors at low charge noise strengths. On the other hand, in the absence of exchange ratio optimization, the insertion of an echo pulse will generally lead to a slight increase in leakage as discussed above.

Additionally, in Fig.~\ref{fig:2q_vs_sigmaV}(c), we show the SAGE CZ infidelity and leakage as a function of $J_\mathrm{max}$ along the optimal path indicated by the red line in Fig.~\ref{fig:pulseramp_vs_maxJ}. This means that we adjust the ramp time for each value of $J_\mathrm{max}$, while we keep the ratio $J_c/J_\mathrm{max}=J_c/J_0$ fixed. We find that in this case the fidelity remains approximately constant as a function of $J_\mathrm{max}$, showing that high-fidelity SAGE two-qubit gates do not necessarily require a very large $J_\mathrm{max}$. This is because even though a larger $J_\mathrm{max}$ (together with a commensurately increased $J_c$) enables a faster gate, it also increases the effects of single-qubit charge noise. As such, there is a trade-off between gate time and charge noise, leading to the infidelity being approximately constant with $J_\mathrm{max}$ here.

Finally, we compare the SAGE CZ infidelity (with an echo pulse) to the infidelity of the Fong-Wandzura CZ (FWCZ) gate for conventional three-dot EO qubits as defined in Ref.~\cite{weinstein_universal_2023}, which consists of a sequence of 26 pulses between nearest-neighbor dots. Specifically, in Fig.~\ref{fig:2q_vs_sigmaV}(b), we plot the infidelities as a function of charge noise strength, showing that SAGE qubits outperform conventional EO qubits up to moderate charge noise strengths due to their enhanced robustness against local magnetic noise. Charge noise does not have any significant effect on the EO FWCZ infidelity, which is limited by magnetic-gradient-induced leakage throughout the entire parameter range we consider. Since these leakage errors cannot be echoed out without dramatically increasing the gate time~\cite{Setiawan2014}, we do not use any echo pulses for the FWCZ gate.

In Fig.~\ref{fig:2q_vs_sigmaV}(d), we further show the SAGE CZ and EO FWCZ infidelities for a fixed charge noise strength as a function of the maximal exchange coupling $J_\mathrm{max}$ that is achievable in a given experimental system. For the SAGE qubit, $J_\mathrm{max}$ sets the always-on coupling $J_0$, and we adjust the pulse ramp for each $J_\mathrm{max}$ according to the red line in Fig.~\ref{fig:pulseramp_vs_maxJ}. For the conventional EO qubit, on the other hand, $J_\mathrm{max}$ sets the pulse strength of the individual pulses in the Fong-Wandzura pulse sequence. As discussed in the context of Fig.~\ref{fig:2q_vs_sigmaV}(c), the SAGE CZ fidelity remains approximately constant here due to a trade-off between gate speed and single-qubit charge noise on the always-on intraqubit couplings, whereas we find that the FWCZ fidelity increases (infidelity decreases) as a function of $J_\mathrm{max}$ as larger pulse amplitudes lead to a faster gate.

\section{Conclusion}
\label{sec:conclusions}

Using a Hubbard model description of the coupled quantum dots, we have demonstrated that the deleterious effects of time-dependent $1/f$ charge noise on the performance of singlet-only always-on gapless exchange (SAGE) spin qubits can be mitigated through simple refocusing pulses reminiscent of CPMG. Such pulses extend the SAGE single-qubit coherence times to hundreds of microseconds for a wide range of experimentally relevant charge noise magnitudes. Furthermore, a single refocusing pulse applied to the individual qubits during a two-qubit entangling operation drives fidelities of more than 99\%, with leakage rates superior to those measured in conventional exchange-only qubits. Importantly, these high-fidelity two-qubit gates can be achieved without any correction for the noise in the entangling interaction itself. All our calculations were performed using noisy, finite-width pulses to perform the necessary rotations, showing that despite pulse imperfections, the refocusing is not compromised to the point of introducing harmful overhead; instead, it substantially improves overall qubit performance. Together, these results clarify that dynamical decoupling approaches should meaningfully improve metrics of future experimental SAGE qubits. Given that SAGE qubits intrinsically suppress the detrimental effects of magnetic-gradient noise by construction, our demonstration of dynamical decoupling induced mitigation of charge noise effects in SAGE qubits is an important milestone. We note that, while our numerical results were obtained assuming a strict $1/f$ form of the noise, the dynamical decoupling strategies discussed here are expected to remain effective also for more general power-law PSDs as long as the noise is predominantly slow.

The principal challenges to SAGE two-qubit gates can be addressed simply by extending the ramp time of the entangling pulses. This leads to practically leakage-free performance in the absence of noise, and reduces control requirements by allowing a broad range of entangling pulse strengths $J_c$. The downside is that ramp duration becomes the dominant contributor to operation runtime, especially as the gate is broken into multiple refocusing segments. It is likely that these stringent ramp time requirements could be substantially relaxed through pulse shaping designed to cancel out the channels to the nearest-lying leakage states~\cite{Polat2026,Kanaar2021,RimbachRuss2023}. However, for simplicity, we have restricted ourselves to pulse shapes consisting of plateaus with finite-duration ramps in this work.

SAGE qubits continue to feature a high ``barrier to entry,'' as they require four dots in a noncollinear arrangement, along with simultaneous operation of exchange couplings, which together place substantial demands on device fabrication and control. However, our results suggest that if these engineering limitations can be overcome, and if the fabrication quality is such that charge noise can be kept below 1~$\mu$eV/$\sqrt{\text{Hz}}$ at 1 Hz, the resulting performance gains may justify these architectural complexities.
Indeed, progress in recent years both in noncollinear dot formation and simultaneous control of exchange couplings~\cite{Broz2025,Broz2026} suggests a promising way forward on these challenges. If similar results can be seen in the four-spin singlet-only encoding, the advantages would include greater flexibility in applying global magnetic fields, as well as the ability to use silicon that is not as highly isotopically purified, leading to more robust supply chains supporting the development of a utility-scale quantum computer using electron spins.

\section*{Acknowledgments}
This work was supported by the Laboratory for Physical Sciences.

\appendix

\section{Detuning maps}

\begin{figure*}[tb]
  \centering
  \includegraphics[width=0.8\linewidth]{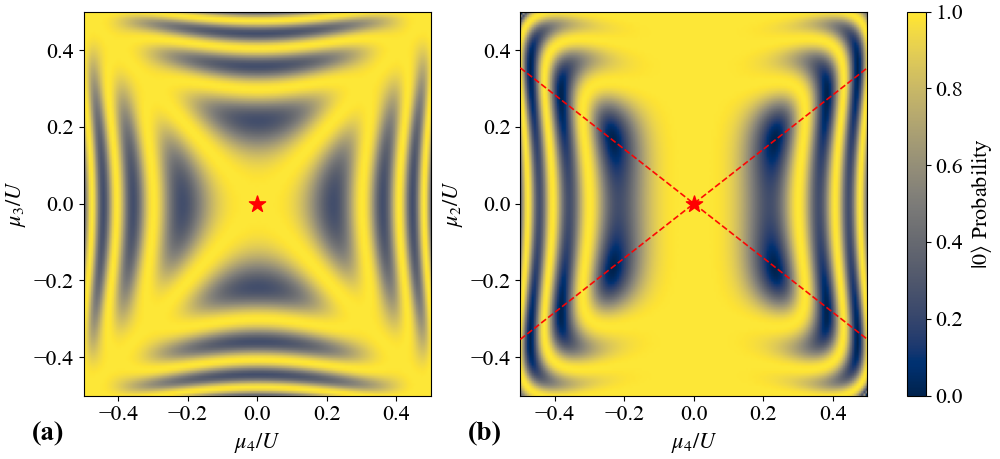}
  \caption{SAGE return probability of measuring the $|0\rangle$ state after pulsing the dot chemical potentials for $\sim\!200$~ns. (a) Return probability as a function of $\mu_3$ and $\mu_4$, while $\mu_1=\mu_2=0$. (b) Return probability as a function of $\mu_2$ and $\mu_4$, while $\mu_1=\mu_3=0$. These return probability maps are consistent with the existence of a full charge noise sweet spot at zero detuning (indicated by a red star). We have set $t_{12}=t_{13}=t_{14}\sim 0.007$~meV and $U=1$~meV in both panels.}
  \label{fig:sweetspot}
\end{figure*}

In Fig.~\ref{fig:sweetspot}, we plot the return probability $P_{|0\rangle}$ after free evolution over a time $t\sim\!200$~ns as a function of the dot chemical potentials. Consistent with the qubit Hamiltonian in Eq.~(\ref{eq:qubit}), we find that rotations around the $z$ axis are realized by detuning the chemical potentials along the diagonal $\mu_3=\mu_4$ in Fig.~\ref{fig:sweetspot}(a) and by detuning $\mu_2$ at fixed $\mu_4=0$ in Fig.~\ref{fig:sweetspot}(b). Return probability maps like the ones shown here could be used to experimentally characterize SAGE spin qubits and, in particular, locate their leakage-protected idle point as a function of dot detunings.

Translations along a single detuning axis do not lead to full population transfer, as the effective exchange being toggled is not in the $x$-$y$ plane and leads to precession about one of the $J$ control axes shown in Fig.~\ref{fig:setup}(d). Only a translation along the path $\frac{1}{2}(J_{14} + J_{13}) = J_{12}$ leads to a rotation about the $x$ axis, leading to a bit-flip of the qubit. For small detuning, Eq.~(\ref{eq:J}) can be expanded as
\begin{equation}
    J_{ij} \approx \frac{4 t_{ij}^2}{U} \left(1 + \left(\frac{\mu_i-\mu_j}{U}\right)^2\right),
\end{equation}
so that when $\mu_1$ is fixed at zero, as is the case in Fig.~\ref{fig:sweetspot}, our condition for $x$ rotations becomes $\frac{1}{2}(\mu_4^2 + \mu_3^2) = \mu_2^2$, which is shown as the dashed red lines in Fig.~\ref{fig:sweetspot}(b), where $\mu_3 = 0$.

\section{Two-qubit adiabatic criterion}
\label{app:adiabatic}

For the two-qubit exchange pulse, we can estimate the ramp times necessary to prevent leakage out of the computational manifold. We parametrize the control by
\begin{equation}
j = \frac{J_{26}}{J_0},
\label{eq:j-ratio}
\end{equation}
with $J_0$ the baseline intraqubit exchange coupling and focus on the interval $0 \le j \le 0.3$. We express the two-qubit Hamiltonian in units of $J_0$, 
\begin{equation}
H(j) = J_0\,\widetilde H(j),
\label{eq:j-physical-hamiltonian}
\end{equation}
where $\widetilde H(j)$ can be expressed in matrix form as
\begin{widetext}
\begin{equation}
\resizebox{\textwidth}{!}{$\displaystyle
\widetilde H(j) =
\left(
\begin{array}{cccccccccccccc}
 -\frac{3}{2} & 0 & 0 & 0 & 0 & 0 & 0 & 0 & -\frac{\sqrt{3} j}{4} & 0 & 0 & 0 & 0 & 0 \\
 0 & -\frac{3}{2} & 0 & 0 & \frac{j}{\sqrt{6}} & 0 & 0 & 0 & 0 & \frac{j}{4 \sqrt{3}} & 0 & 0 & 0 & 0 \\
 0 & 0 & -\frac{3}{2} & 0 & 0 & 0 & \frac{j}{\sqrt{6}} & 0 & 0 & 0 & \frac{j}{4 \sqrt{3}} & 0 & 0 & 0 \\
 0 & 0 & 0 & -\frac{3}{2} & 0 & -\frac{j}{3 \sqrt{6}} & 0 & -\frac{j}{3 \sqrt{6}} & 0 & 0 & 0 & -\frac{j}{12
   \sqrt{3}} & -\frac{2 j}{3 \sqrt{3}} & 0 \\
 0 & \frac{j}{\sqrt{6}} & 0 & 0 & -\frac{5 j}{12}-1 & 0 & 0 & 0 & 0 & -\frac{j}{3 \sqrt{2}} & 0 & 0 & 0 & 0 \\
 0 & 0 & 0 & -\frac{j}{3 \sqrt{6}} & 0 & \frac{5 j}{36}-1 & 0 & -\frac{j}{9} & 0 & 0 & 0 & \frac{j}{9
   \sqrt{2}} & -\frac{5 j}{18 \sqrt{2}} & \frac{1}{6} \sqrt{\frac{5}{6}} j \\
 0 & 0 & \frac{j}{\sqrt{6}} & 0 & 0 & 0 & -\frac{5 j}{12}-1 & 0 & 0 & 0 & -\frac{j}{3 \sqrt{2}} & 0 & 0 & 0 \\
 0 & 0 & 0 & -\frac{j}{3 \sqrt{6}} & 0 & -\frac{j}{9} & 0 & \frac{5 j}{36}-1 & 0 & 0 & 0 & \frac{j}{9
   \sqrt{2}} & -\frac{5 j}{18 \sqrt{2}} & \frac{1}{6} \sqrt{\frac{5}{6}} j \\
 -\frac{\sqrt{3} j}{4} & 0 & 0 & 0 & 0 & 0 & 0 & 0 & \frac{1}{2}-\frac{j}{2} & 0 & 0 & 0 & 0 & 0 \\
 0 & \frac{j}{4 \sqrt{3}} & 0 & 0 & -\frac{j}{3 \sqrt{2}} & 0 & 0 & 0 & 0 & \frac{j}{6}+\frac{1}{2} & 0 & 0 &
   0 & 0 \\
 0 & 0 & \frac{j}{4 \sqrt{3}} & 0 & 0 & 0 & -\frac{j}{3 \sqrt{2}} & 0 & 0 & 0 & \frac{j}{6}+\frac{1}{2} & 0 &
   0 & 0 \\
 0 & 0 & 0 & -\frac{j}{12 \sqrt{3}} & 0 & \frac{j}{9 \sqrt{2}} & 0 & \frac{j}{9 \sqrt{2}} & 0 & 0 & 0 &
   \frac{1}{2}-\frac{j}{18} & -\frac{j}{9} & -\frac{1}{3} \sqrt{\frac{5}{3}} j \\
 0 & 0 & 0 & -\frac{2 j}{3 \sqrt{3}} & 0 & -\frac{5 j}{18 \sqrt{2}} & 0 & -\frac{5 j}{18 \sqrt{2}} & 0 & 0 & 0
   & -\frac{j}{9} & -\frac{25 j}{72}-\frac{5}{2} & -\frac{1}{24} \sqrt{\frac{5}{3}} j \\
 0 & 0 & 0 & 0 & 0 & \frac{1}{6} \sqrt{\frac{5}{6}} j & 0 & \frac{1}{6} \sqrt{\frac{5}{6}} j & 0 & 0 & 0 &
   -\frac{1}{3} \sqrt{\frac{5}{3}} j & -\frac{1}{24} \sqrt{\frac{5}{3}} j & \frac{3}{2}-\frac{3 j}{8} \\
\end{array}
\right)
$},
\label{eq:j-dimensionless-hamiltonian}
\end{equation}
\end{widetext} 
using the basis states given in Table~\ref{tab:pq-quantum-numbers} ($P$ denotes a computational state, and $Q$ a leakage state). Although on the computational subspace it is equally valid and perhaps more natural to use the intraqubit quantum numbers $S_{12}$ and $S_{34}$, we must work with the quantum numbers $s_{34},S_{234}$ on the left and $s_{78},S_{678}$ on the right to diagonalize the full $S=0$ Hilbert space including leakage states.

The adiabatic criterion between states $i$ and $k$ can be expressed as
\begin{equation}
\left|\frac{dj}{dt}\right| \chi_{ik}(j) \ll \frac{1}{\hbar},
\;\;\;
\chi_{ik}(j) =
\frac{\left|\left\langle Q_k(j) \middle| \partial_j H \middle| P_i(j) \right\rangle\right|}{\left[E_{Q_k}(j) - E_{P_i}(j)\right]^2}.
\label{eq:j-susceptibility}
\end{equation}
This inequality must be satisfied between all pairings at all times in the simulation in order to prevent intrinsic leakage out of the computational subspace.

The nearest-lying leakage states are $Q_{5-8}$ and between this space and the computational space, the pairs $P_2$ and $Q_5$ and, by symmetry, $P_3$ and $Q_7$, feature the largest couplings. Analyzing this representative pair of states allows us to estimate the $j$ ramp times needed to quell leakage.

\begin{figure}[tb]
  \centering
  \includegraphics[width=\columnwidth]{j26_spectrum.png}
  \caption{Relevant eigenenergies of $H(j)$ for $0 \le j \le 0.3$. Solid curves denote the computational states $P_1,\ldots,P_4$, while dot-dash curves denote the nearest energy leakage states $Q_5,\ldots,Q_8$. Energies are shown in units of $J_0$.}
  \label{fig:j26-spectrum}
\end{figure}

Throughout the ramp of $j$, the minimal energy gap (see Fig.~\ref{fig:j26-spectrum}) is 
\begin{equation}
    \Delta_{P_2Q_5} \approx 0.45 J_0,
\end{equation}
while the corresponding bare matrix element is
\begin{equation}
\left\langle Q_5 \middle| \partial_j (J_0 \widetilde H) \middle| P_2 \right\rangle = \frac{J_0}{\sqrt{6}}.
\end{equation}
This gives the estimate
\begin{equation}
\chi_{\max} \approx \frac{J_0/\sqrt{6}}{(0.45 J_0)^2}
\approx \frac{2}{J_0},
\end{equation}
so that for a linear ramp one obtains
\begin{equation}
\left|\frac{dj}{dt}\right| \ll \frac{1}{\hbar \chi_{\max}}
\approx \frac{J_0}{2\hbar}.
\end{equation}
If $j$ is ramped from $0$ to $j_f$ in a total time $\tau$, so that
\begin{equation}
\left|\frac{dj}{dt}\right|=\frac{j_f}{\tau},
\end{equation}
then adiabaticity requires
\begin{equation}
\tau \gg \frac{2\hbar}{J_0}\, j_f \equiv \tau_0.
\label{eq:j-ramp-time}
\end{equation}

As a concrete example, for $j_f = 0.2$ and $J_0/h = 50~\mathrm{MHz}$, one has $J_0/\hbar = 2\pi \times 50~\mathrm{MHz} \approx 3.14 \times 10^8~\mathrm{s}^{-1}$. The adiabatic condition then becomes $\tau \gg 2j_f/(J_0/\hbar) \approx 1.3~\mathrm{ns}$. Empirically, we have found that $\tau > 50 \tau_0$ is usually sufficient to generally suppress leakage as a major source of error (for example, see Fig.~\ref{fig:pulseramp_vs_maxJ}).

\begin{table*}[htb]
  \centering
  \small
  \begin{tabular}{c c c c c c c c c c c c c c c}
    \hline\hline
    & $P_1$ & $P_2$ & $P_3$ & $P_4$ & $Q_5$ & $Q_6$ & $Q_7$ & $Q_8$ & $Q_9$ & $Q_{10}$ & $Q_{11}$ & $Q_{12}$ & $Q_{13}$ & $Q_{14}$ \\
    \hline
    $S_{1234}$ & $0$ & $0$ & $0$ & $0$ & $1$ & $1$ & $1$ & $1$ & $1$ & $1$ & $1$ & $1$ & $1$ & $2$ \\
    $S_{5678}$ & $0$ & $0$ & $0$ & $0$ & $1$ & $1$ & $1$ & $1$ & $1$ & $1$ & $1$ & $1$ & $1$ & $2$ \\
    $S_{234}$ & $\frac{1}{2}$ & $\frac{1}{2}$ & $\frac{1}{2}$ & $\frac{1}{2}$ & $\frac{1}{2}$ & $\frac{1}{2}$ & $\frac{3}{2}$ & $\frac{3}{2}$ & $\frac{1}{2}$ & $\frac{1}{2}$ & $\frac{1}{2}$ & $\frac{1}{2}$ & $\frac{3}{2}$ & $\frac{3}{2}$ \\
    $S_{678}$ & $\frac{1}{2}$ & $\frac{1}{2}$ & $\frac{1}{2}$ & $\frac{1}{2}$ & $\frac{3}{2}$ & $\frac{3}{2}$ & $\frac{1}{2}$ & $\frac{1}{2}$ & $\frac{1}{2}$ & $\frac{1}{2}$ & $\frac{1}{2}$ & $\frac{1}{2}$ & $\frac{3}{2}$ & $\frac{3}{2}$ \\
    $s_{34}$ & $0$ & $0$ & $1$ & $1$ & $0$ & $1$ & $1$ & $1$ & $0$ & $0$ & $1$ & $1$ & $1$ & $1$ \\
    $s_{78}$ & $0$ & $1$ & $0$ & $1$ & $1$ & $1$ & $0$ & $1$ & $0$ & $1$ & $0$ & $1$ & $1$ & $1$ \\
    \hline\hline
  \end{tabular}
  \caption{Quantum numbers for the 14-state basis with total $S=0$.}
  \label{tab:pq-quantum-numbers}
\end{table*}

\section{Coherent leakage oscillations}
\label{app:oscillations}

As established in Appendix~\ref{app:adiabatic}, the dominant leakage channels for this SAGE two-qubit system are between $P_2$ and $Q_5$ and, by symmetry, between $P_3$ and $Q_7$. These pairs form two identical two-level subspaces:
\begin{equation}H_\text{proj}  = \sum_{k,l\in\{2,5\}} H_{kl} |k\rangle\langle l|= \begin{pmatrix}-\frac{3}{2}J_0 & \frac{1}{\sqrt{6}}J_{26} \\ \frac{1}{\sqrt{6}}J_{26} & -\frac{5}{12}J_{26} - J_0\end{pmatrix}. \end{equation}
After subtracting the trace contribution, we have
\begin{equation}H_\text{proj} = \begin{pmatrix}-\frac{1}{4}J_0 + \frac{5}{24}J_{26} & \frac{1}{\sqrt{6}}J_{26} \\ \frac{1}{\sqrt{6}}J_{26} & \frac{1}{4}J_0 - \frac{5}{24}J_{26} \end{pmatrix}, \end{equation}
which leads to the familiar dressed state dynamics:
\begin{equation} H = \sqrt{\Delta^2 + \gamma^2} \begin{pmatrix} \cos\theta & \sin\theta \\ \sin\theta  & -\cos\theta \end{pmatrix}, \end{equation}
with eigenenergies $E_\pm = \pm\sqrt{\Delta^2 + \gamma^2}$ and the states now dressed with a skew angle,
\begin{equation} \tan \theta = \frac{\gamma}{\Delta},\quad\Delta = E_+ \cos\theta,\quad\gamma = E_+ \sin\theta,\end{equation}
where
\begin{equation}\Delta = -\frac{1}{4}J_0 +\frac{5}{24}J_{26} \quad\text{and}\quad \gamma = J_{26}/\sqrt{6}.\end{equation}
The eigenstates of this Hamiltonian are
\begin{align} \ket{+} &= \cos \frac{\theta}{2} \ket{0} + \sin\frac{\theta}{2} \ket{1}, \\ \ket{-} &= \sin \frac{\theta}{2} \ket{0} - \cos\frac{\theta}{2} \ket{1},\end{align}
and one can write down the leakage probability as a function of time, which is
\begin{align}
P_{0\rightarrow1} &= |\langle1|e^{-i H t/\hbar}|0\rangle|^2  =  \sin^2\theta\sin^2\left(\frac{t\sqrt{\Delta^2 + \gamma^2} }{\hbar}\right) \\ \label{eq:dressed}&=\frac{\gamma^2}{\Delta^2 + \gamma^2} \sin^2\left(\frac{t\sqrt{\Delta^2 + \gamma^2} }{\hbar}\right).
\end{align}
SAGE two-qubit gate durations are determined by the integral $\int J_\text{eff}(t)/\hbar\; dt = \pi/4$, with the entangling strength and gate time (assuming square pulses)
\begin{equation} 
J_\text{eff} \approx \frac{J_{26}^2}{6J_0}  + \frac{5J_{26}^3}{32J_0^2};\quad t = \frac{\pi\hbar}{4J_\text{eff}}.
\end{equation}
Substituting into Eq.~(\ref{eq:dressed}) yields
\begin{equation}
P_{\text{leak}}^{\text{dom}}= \frac{\frac{8}{3}j^2}{1 - \frac{5}{3}j + \frac{121}{36}j^2 } \sin^2 \left(\frac{\pi\sqrt{1 - \frac{5}{3}j + \frac{121}{36}j^2}}{ \frac{8}{3}j^2  + \frac{5}{2}j^3}\right).
\end{equation}
The leakage oscillations in Figs.~\ref{fig:ramptime}(a) and \ref{fig:ramptime}(b) are qualitatively described by this expression.

For the refocused sequence [see Fig.~\ref{fig:ramptime}(c)], the interqubit pulse is divided into two equal halves, which halves the argument of the sinusoid accumulated during each pulse segment. This is why the choice of $J_c/J_0 \approx 0.2$, which minimizes the leakage for the unrefocused pulse, actually maximizes leakage in the case of an echo. Here, the argument of the sinusoid comes out to approximately $ 7 \pi$ for the unrefocused pulse. A better choice for $j$ would lead to an even multiple of $\pi$, preferably a power of 2 for the purpose of repeated refocusing.  A choice such as $j \approx 0.189$, for which the argument of the sinusoid is approximately $8\pi$, combined with pulse optimization to reduce needed ramp times could lead to further fidelity improvements for refocused SAGE two-qubit gates.

\bibliography{references4}

\begin{thebibliography}{45}%
\makeatletter
\providecommand \@ifxundefined [1]{%
 \@ifx{#1\undefined}
}%
\providecommand \@ifnum [1]{%
 \ifnum #1\expandafter \@firstoftwo
 \else \expandafter \@secondoftwo
 \fi
}%
\providecommand \@ifx [1]{%
 \ifx #1\expandafter \@firstoftwo
 \else \expandafter \@secondoftwo
 \fi
}%
\providecommand \natexlab [1]{#1}%
\providecommand \enquote  [1]{``#1''}%
\providecommand \bibnamefont  [1]{#1}%
\providecommand \bibfnamefont [1]{#1}%
\providecommand \citenamefont [1]{#1}%
\providecommand \href@noop [0]{\@secondoftwo}%
\providecommand \href [0]{\begingroup \@sanitize@url \@href}%
\providecommand \@href[1]{\@@startlink{#1}\@@href}%
\providecommand \@@href[1]{\endgroup#1\@@endlink}%
\providecommand \@sanitize@url [0]{\catcode `\\12\catcode `\$12\catcode
  `\&12\catcode `\#12\catcode `\^12\catcode `\_12\catcode `\%12\relax}%
\providecommand \@@startlink[1]{}%
\providecommand \@@endlink[0]{}%
\providecommand \url  [0]{\begingroup\@sanitize@url \@url }%
\providecommand \@url [1]{\endgroup\@href {#1}{\urlprefix }}%
\providecommand \urlprefix  [0]{URL }%
\providecommand \Eprint [0]{\href }%
\providecommand \doibase [0]{https://doi.org/}%
\providecommand \selectlanguage [0]{\@gobble}%
\providecommand \bibinfo  [0]{\@secondoftwo}%
\providecommand \bibfield  [0]{\@secondoftwo}%
\providecommand \translation [1]{[#1]}%
\providecommand \BibitemOpen [0]{}%
\providecommand \bibitemStop [0]{}%
\providecommand \bibitemNoStop [0]{.\EOS\space}%
\providecommand \EOS [0]{\spacefactor3000\relax}%
\providecommand \BibitemShut  [1]{\csname bibitem#1\endcsname}%
\let\auto@bib@innerbib\@empty
%</preamble>
\bibitem [{\citenamefont {Andrews}\ \emph {et~al.}(2019)\citenamefont
  {Andrews}, \citenamefont {Jones}, \citenamefont {Reed}, \citenamefont
  {Jones}, \citenamefont {Ha}, \citenamefont {Jura}, \citenamefont {Kerckhoff},
  \citenamefont {Levendorf}, \citenamefont {Meenehan}, \citenamefont {Merkel},
  \citenamefont {Smith}, \citenamefont {Sun}, \citenamefont {Weinstein},
  \citenamefont {Rakher}, \citenamefont {Ladd},\ and\ \citenamefont
  {Borselli}}]{andrews_quantifying_2019}%
  \BibitemOpen
  \bibfield  {author} {\bibinfo {author} {\bibfnamefont {R.~W.}\ \bibnamefont
  {Andrews}}, \bibinfo {author} {\bibfnamefont {C.}~\bibnamefont {Jones}},
  \bibinfo {author} {\bibfnamefont {M.~D.}\ \bibnamefont {Reed}}, \bibinfo
  {author} {\bibfnamefont {A.~M.}\ \bibnamefont {Jones}}, \bibinfo {author}
  {\bibfnamefont {S.~D.}\ \bibnamefont {Ha}}, \bibinfo {author} {\bibfnamefont
  {M.~P.}\ \bibnamefont {Jura}}, \bibinfo {author} {\bibfnamefont
  {J.}~\bibnamefont {Kerckhoff}}, \bibinfo {author} {\bibfnamefont
  {M.}~\bibnamefont {Levendorf}}, \bibinfo {author} {\bibfnamefont
  {S.}~\bibnamefont {Meenehan}}, \bibinfo {author} {\bibfnamefont {S.~T.}\
  \bibnamefont {Merkel}}, \bibinfo {author} {\bibfnamefont {A.}~\bibnamefont
  {Smith}}, \bibinfo {author} {\bibfnamefont {B.}~\bibnamefont {Sun}}, \bibinfo
  {author} {\bibfnamefont {A.~J.}\ \bibnamefont {Weinstein}}, \bibinfo {author}
  {\bibfnamefont {M.~T.}\ \bibnamefont {Rakher}}, \bibinfo {author}
  {\bibfnamefont {T.~D.}\ \bibnamefont {Ladd}},\ and\ \bibinfo {author}
  {\bibfnamefont {M.~G.}\ \bibnamefont {Borselli}},\ }\bibfield  {title}
  {\bibinfo {title} {Quantifying error and leakage in an encoded {Si/SiGe}
  triple-dot qubit},\ }\href {https://doi.org/10.1038/s41565-019-0500-4}
  {\bibfield  {journal} {\bibinfo  {journal} {Nat. Nanotechnol.}\ }\textbf
  {\bibinfo {volume} {14}},\ \bibinfo {pages} {747} (\bibinfo {year}
  {2019})}\BibitemShut {NoStop}%
\bibitem [{\citenamefont {Kerckhoff}\ \emph {et~al.}(2021)\citenamefont
  {Kerckhoff}, \citenamefont {Sun}, \citenamefont {Fong}, \citenamefont
  {Jones}, \citenamefont {Kiselev}, \citenamefont {Barnes}, \citenamefont
  {Noah}, \citenamefont {Acuna}, \citenamefont {Akmal}, \citenamefont {Ha},
  \citenamefont {Wright}, \citenamefont {Thomas}, \citenamefont {Jackson},
  \citenamefont {Edge}, \citenamefont {Eng}, \citenamefont {Ross},\ and\
  \citenamefont {Ladd}}]{kerckhoff_magnetic_2021}%
  \BibitemOpen
  \bibfield  {author} {\bibinfo {author} {\bibfnamefont {J.}~\bibnamefont
  {Kerckhoff}}, \bibinfo {author} {\bibfnamefont {B.}~\bibnamefont {Sun}},
  \bibinfo {author} {\bibfnamefont {B.}~\bibnamefont {Fong}}, \bibinfo {author}
  {\bibfnamefont {C.}~\bibnamefont {Jones}}, \bibinfo {author} {\bibfnamefont
  {A.}~\bibnamefont {Kiselev}}, \bibinfo {author} {\bibfnamefont
  {D.}~\bibnamefont {Barnes}}, \bibinfo {author} {\bibfnamefont
  {R.}~\bibnamefont {Noah}}, \bibinfo {author} {\bibfnamefont {E.}~\bibnamefont
  {Acuna}}, \bibinfo {author} {\bibfnamefont {M.}~\bibnamefont {Akmal}},
  \bibinfo {author} {\bibfnamefont {S.}~\bibnamefont {Ha}}, \bibinfo {author}
  {\bibfnamefont {J.}~\bibnamefont {Wright}}, \bibinfo {author} {\bibfnamefont
  {B.}~\bibnamefont {Thomas}}, \bibinfo {author} {\bibfnamefont
  {C.}~\bibnamefont {Jackson}}, \bibinfo {author} {\bibfnamefont
  {L.}~\bibnamefont {Edge}}, \bibinfo {author} {\bibfnamefont {K.}~\bibnamefont
  {Eng}}, \bibinfo {author} {\bibfnamefont {R.}~\bibnamefont {Ross}},\ and\
  \bibinfo {author} {\bibfnamefont {T.}~\bibnamefont {Ladd}},\ }\bibfield
  {title} {\bibinfo {title} {Magnetic gradient fluctuations from quadrupolar
  ${}^{73}\mathrm{Ge}$ in $\mathrm{Si}$/$\mathrm{Si}\mathrm{Ge}$ exchange-only
  qubits},\ }\href {https://doi.org/10.1103/PRXQuantum.2.010347} {\bibfield
  {journal} {\bibinfo  {journal} {PRX Quantum}\ }\textbf {\bibinfo {volume}
  {2}},\ \bibinfo {pages} {010347} (\bibinfo {year} {2021})}\BibitemShut
  {NoStop}%
\bibitem [{\citenamefont {Ha}\ \emph {et~al.}(2022)\citenamefont {Ha},
  \citenamefont {Ha}, \citenamefont {Choi}, \citenamefont {Tang}, \citenamefont
  {Schmitz}, \citenamefont {Levendorf}, \citenamefont {Lee}, \citenamefont
  {Chappell}, \citenamefont {Adams}, \citenamefont {Hulbert}, \citenamefont
  {Acuna}, \citenamefont {Noah}, \citenamefont {Matten}, \citenamefont {Jura},
  \citenamefont {Wright}, \citenamefont {Rakher},\ and\ \citenamefont
  {Borselli}}]{ha_flexible_2022}%
  \BibitemOpen
  \bibfield  {author} {\bibinfo {author} {\bibfnamefont {W.}~\bibnamefont
  {Ha}}, \bibinfo {author} {\bibfnamefont {S.~D.}\ \bibnamefont {Ha}}, \bibinfo
  {author} {\bibfnamefont {M.~D.}\ \bibnamefont {Choi}}, \bibinfo {author}
  {\bibfnamefont {Y.}~\bibnamefont {Tang}}, \bibinfo {author} {\bibfnamefont
  {A.~E.}\ \bibnamefont {Schmitz}}, \bibinfo {author} {\bibfnamefont {M.~P.}\
  \bibnamefont {Levendorf}}, \bibinfo {author} {\bibfnamefont {K.}~\bibnamefont
  {Lee}}, \bibinfo {author} {\bibfnamefont {J.~M.}\ \bibnamefont {Chappell}},
  \bibinfo {author} {\bibfnamefont {T.~S.}\ \bibnamefont {Adams}}, \bibinfo
  {author} {\bibfnamefont {D.~R.}\ \bibnamefont {Hulbert}}, \bibinfo {author}
  {\bibfnamefont {E.}~\bibnamefont {Acuna}}, \bibinfo {author} {\bibfnamefont
  {R.~S.}\ \bibnamefont {Noah}}, \bibinfo {author} {\bibfnamefont {J.~W.}\
  \bibnamefont {Matten}}, \bibinfo {author} {\bibfnamefont {M.~P.}\
  \bibnamefont {Jura}}, \bibinfo {author} {\bibfnamefont {J.~A.}\ \bibnamefont
  {Wright}}, \bibinfo {author} {\bibfnamefont {M.~T.}\ \bibnamefont {Rakher}},\
  and\ \bibinfo {author} {\bibfnamefont {M.~G.}\ \bibnamefont {Borselli}},\
  }\bibfield  {title} {\bibinfo {title} {A {Flexible} {Design} {Platform} for
  {Si}/{SiGe} {Exchange}-{Only} {Qubits} with {Low} {Disorder}},\ }\href
  {https://doi.org/10.1021/acs.nanolett.1c03026} {\bibfield  {journal}
  {\bibinfo  {journal} {Nano Letters}\ }\textbf {\bibinfo {volume} {22}},\
  \bibinfo {pages} {1443} (\bibinfo {year} {2022})}\BibitemShut {NoStop}%
\bibitem [{\citenamefont {Weinstein}\ \emph {et~al.}(2023)\citenamefont
  {Weinstein}, \citenamefont {Reed}, \citenamefont {Jones}, \citenamefont
  {Andrews}, \citenamefont {Barnes}, \citenamefont {Blumoff}, \citenamefont
  {Euliss}, \citenamefont {Eng}, \citenamefont {Fong}, \citenamefont {Ha},
  \citenamefont {Hulbert}, \citenamefont {Jackson}, \citenamefont {Jura},
  \citenamefont {Keating}, \citenamefont {Kerckhoff}, \citenamefont {Kiselev},
  \citenamefont {Matten}, \citenamefont {Sabbir}, \citenamefont {Smith},
  \citenamefont {Wright}, \citenamefont {Rakher}, \citenamefont {Ladd},\ and\
  \citenamefont {Borselli}}]{weinstein_universal_2023}%
  \BibitemOpen
  \bibfield  {author} {\bibinfo {author} {\bibfnamefont {A.}~\bibnamefont
  {Weinstein}}, \bibinfo {author} {\bibfnamefont {M.}~\bibnamefont {Reed}},
  \bibinfo {author} {\bibfnamefont {A.}~\bibnamefont {Jones}}, \bibinfo
  {author} {\bibfnamefont {R.}~\bibnamefont {Andrews}}, \bibinfo {author}
  {\bibfnamefont {D.}~\bibnamefont {Barnes}}, \bibinfo {author} {\bibfnamefont
  {J.}~\bibnamefont {Blumoff}}, \bibinfo {author} {\bibfnamefont
  {L.}~\bibnamefont {Euliss}}, \bibinfo {author} {\bibfnamefont
  {K.}~\bibnamefont {Eng}}, \bibinfo {author} {\bibfnamefont {B.}~\bibnamefont
  {Fong}}, \bibinfo {author} {\bibfnamefont {S.}~\bibnamefont {Ha}}, \bibinfo
  {author} {\bibfnamefont {D.}~\bibnamefont {Hulbert}}, \bibinfo {author}
  {\bibfnamefont {C.}~\bibnamefont {Jackson}}, \bibinfo {author} {\bibfnamefont
  {M.}~\bibnamefont {Jura}}, \bibinfo {author} {\bibfnamefont {T.}~\bibnamefont
  {Keating}}, \bibinfo {author} {\bibfnamefont {J.}~\bibnamefont {Kerckhoff}},
  \bibinfo {author} {\bibfnamefont {A.}~\bibnamefont {Kiselev}}, \bibinfo
  {author} {\bibfnamefont {J.}~\bibnamefont {Matten}}, \bibinfo {author}
  {\bibfnamefont {G.}~\bibnamefont {Sabbir}}, \bibinfo {author} {\bibfnamefont
  {A.}~\bibnamefont {Smith}}, \bibinfo {author} {\bibfnamefont
  {J.}~\bibnamefont {Wright}}, \bibinfo {author} {\bibfnamefont
  {M.}~\bibnamefont {Rakher}}, \bibinfo {author} {\bibfnamefont
  {T.}~\bibnamefont {Ladd}},\ and\ \bibinfo {author} {\bibfnamefont
  {M.}~\bibnamefont {Borselli}},\ }\bibfield  {title} {\bibinfo {title}
  {Universal logic with encoded spin qubits in silicon},\ }\href
  {https://www.nature.com/articles/s41586-023-05777-3} {\bibfield  {journal}
  {\bibinfo  {journal} {Nature}\ }\textbf {\bibinfo {volume} {615}},\ \bibinfo
  {pages} {817} (\bibinfo {year} {2023})}\BibitemShut {NoStop}%
\bibitem [{\citenamefont {Heinz}\ \emph {et~al.}(2025)\citenamefont {Heinz},
  \citenamefont {Borjans}, \citenamefont {Curry}, \citenamefont {Kotlyar},
  \citenamefont {Luthi}, \citenamefont {M\k{a}dzik}, \citenamefont
  {Mohiyaddin}, \citenamefont {Bishop},\ and\ \citenamefont
  {Burkard}}]{heinz_fast_2024}%
  \BibitemOpen
  \bibfield  {author} {\bibinfo {author} {\bibfnamefont {I.}~\bibnamefont
  {Heinz}}, \bibinfo {author} {\bibfnamefont {F.}~\bibnamefont {Borjans}},
  \bibinfo {author} {\bibfnamefont {M.~J.}\ \bibnamefont {Curry}}, \bibinfo
  {author} {\bibfnamefont {R.}~\bibnamefont {Kotlyar}}, \bibinfo {author}
  {\bibfnamefont {F.}~\bibnamefont {Luthi}}, \bibinfo {author} {\bibfnamefont
  {M.~T.}\ \bibnamefont {M\k{a}dzik}}, \bibinfo {author} {\bibfnamefont
  {F.~A.}\ \bibnamefont {Mohiyaddin}}, \bibinfo {author} {\bibfnamefont
  {N.}~\bibnamefont {Bishop}},\ and\ \bibinfo {author} {\bibfnamefont
  {G.}~\bibnamefont {Burkard}},\ }\bibfield  {title} {\bibinfo {title} {Fast
  quantum gates for exchange-only qubits using simultaneous exchange pulses},\
  }\href {https://doi.org/10.1103/njq3-fcdd} {\bibfield  {journal} {\bibinfo
  {journal} {PRX Quantum}\ }\textbf {\bibinfo {volume} {6}},\ \bibinfo {pages}
  {030353} (\bibinfo {year} {2025})}\BibitemShut {NoStop}%
\bibitem [{\citenamefont {Acuna}\ \emph {et~al.}(2024)\citenamefont {Acuna},
  \citenamefont {Broz}, \citenamefont {Shyamsundar}, \citenamefont {Mei},
  \citenamefont {Feeney}, \citenamefont {Smetanka}, \citenamefont {Davis},
  \citenamefont {Lee}, \citenamefont {Choi}, \citenamefont {Boyd},
  \citenamefont {Suh}, \citenamefont {Ha}, \citenamefont {Jennings},
  \citenamefont {Pan}, \citenamefont {Sanchez}, \citenamefont {Reed},\ and\
  \citenamefont {Petta}}]{acuna_coherent_2024}%
  \BibitemOpen
  \bibfield  {author} {\bibinfo {author} {\bibfnamefont {E.}~\bibnamefont
  {Acuna}}, \bibinfo {author} {\bibfnamefont {J.~D.}\ \bibnamefont {Broz}},
  \bibinfo {author} {\bibfnamefont {K.}~\bibnamefont {Shyamsundar}}, \bibinfo
  {author} {\bibfnamefont {A.~B.}\ \bibnamefont {Mei}}, \bibinfo {author}
  {\bibfnamefont {C.~P.}\ \bibnamefont {Feeney}}, \bibinfo {author}
  {\bibfnamefont {V.}~\bibnamefont {Smetanka}}, \bibinfo {author}
  {\bibfnamefont {T.}~\bibnamefont {Davis}}, \bibinfo {author} {\bibfnamefont
  {K.}~\bibnamefont {Lee}}, \bibinfo {author} {\bibfnamefont {M.~D.}\
  \bibnamefont {Choi}}, \bibinfo {author} {\bibfnamefont {B.}~\bibnamefont
  {Boyd}}, \bibinfo {author} {\bibfnamefont {J.}~\bibnamefont {Suh}}, \bibinfo
  {author} {\bibfnamefont {W.}~\bibnamefont {Ha}}, \bibinfo {author}
  {\bibfnamefont {C.}~\bibnamefont {Jennings}}, \bibinfo {author}
  {\bibfnamefont {A.~S.}\ \bibnamefont {Pan}}, \bibinfo {author} {\bibfnamefont
  {D.~S.}\ \bibnamefont {Sanchez}}, \bibinfo {author} {\bibfnamefont {M.~D.}\
  \bibnamefont {Reed}},\ and\ \bibinfo {author} {\bibfnamefont {J.~R.}\
  \bibnamefont {Petta}},\ }\bibfield  {title} {\bibinfo {title} {Coherent
  control of a triangular exchange-only spin qubit},\ }\href
  {https://link.aps.org/doi/10.1103/PhysRevApplied.22.044057} {\bibfield
  {journal} {\bibinfo  {journal} {Phys. Rev. Appl.}\ }\textbf {\bibinfo
  {volume} {22}},\ \bibinfo {pages} {044057} (\bibinfo {year}
  {2024})}\BibitemShut {NoStop}%
\bibitem [{\citenamefont {Mądzik}\ \emph {et~al.}(2025)\citenamefont
  {Mądzik}, \citenamefont {Luthi}, \citenamefont {Guerreschi}, \citenamefont
  {Mohiyaddin}, \citenamefont {Borjans}, \citenamefont {Chadwick},
  \citenamefont {Curry}, \citenamefont {Ziegler}, \citenamefont {Atanasov},
  \citenamefont {Bavdaz}, \citenamefont {Connors}, \citenamefont {Corrigan},
  \citenamefont {Ercan}, \citenamefont {Flory}, \citenamefont {George},
  \citenamefont {Harpt}, \citenamefont {Henry}, \citenamefont {Islam},
  \citenamefont {Khammassi}, \citenamefont {Keith}, \citenamefont {Lampert},
  \citenamefont {Mladenov}, \citenamefont {Morris}, \citenamefont {Nethwewala},
  \citenamefont {Neyens}, \citenamefont {Otten}, \citenamefont {Osuna~Ibarra},
  \citenamefont {Patra}, \citenamefont {Pillarisetty}, \citenamefont
  {Premaratne}, \citenamefont {Ramsey}, \citenamefont {Risinger}, \citenamefont
  {Rooney}, \citenamefont {Savytskyy}, \citenamefont {Watson}, \citenamefont
  {Zietz}, \citenamefont {Matsuura}, \citenamefont {Pellerano}, \citenamefont
  {Bishop}, \citenamefont {Roberts},\ and\ \citenamefont
  {Clarke}}]{Madzik2025}%
  \BibitemOpen
  \bibfield  {author} {\bibinfo {author} {\bibfnamefont {M.~T.}\ \bibnamefont
  {Mądzik}}, \bibinfo {author} {\bibfnamefont {F.}~\bibnamefont {Luthi}},
  \bibinfo {author} {\bibfnamefont {G.~G.}\ \bibnamefont {Guerreschi}},
  \bibinfo {author} {\bibfnamefont {F.~A.}\ \bibnamefont {Mohiyaddin}},
  \bibinfo {author} {\bibfnamefont {F.}~\bibnamefont {Borjans}}, \bibinfo
  {author} {\bibfnamefont {J.~D.}\ \bibnamefont {Chadwick}}, \bibinfo {author}
  {\bibfnamefont {M.~J.}\ \bibnamefont {Curry}}, \bibinfo {author}
  {\bibfnamefont {J.}~\bibnamefont {Ziegler}}, \bibinfo {author} {\bibfnamefont
  {S.}~\bibnamefont {Atanasov}}, \bibinfo {author} {\bibfnamefont {P.~L.}\
  \bibnamefont {Bavdaz}}, \bibinfo {author} {\bibfnamefont {E.~J.}\
  \bibnamefont {Connors}}, \bibinfo {author} {\bibfnamefont {J.}~\bibnamefont
  {Corrigan}}, \bibinfo {author} {\bibfnamefont {H.~E.}\ \bibnamefont {Ercan}},
  \bibinfo {author} {\bibfnamefont {R.}~\bibnamefont {Flory}}, \bibinfo
  {author} {\bibfnamefont {H.~C.}\ \bibnamefont {George}}, \bibinfo {author}
  {\bibfnamefont {B.}~\bibnamefont {Harpt}}, \bibinfo {author} {\bibfnamefont
  {E.}~\bibnamefont {Henry}}, \bibinfo {author} {\bibfnamefont {M.~M.}\
  \bibnamefont {Islam}}, \bibinfo {author} {\bibfnamefont {N.}~\bibnamefont
  {Khammassi}}, \bibinfo {author} {\bibfnamefont {D.}~\bibnamefont {Keith}},
  \bibinfo {author} {\bibfnamefont {L.~F.}\ \bibnamefont {Lampert}}, \bibinfo
  {author} {\bibfnamefont {T.~M.}\ \bibnamefont {Mladenov}}, \bibinfo {author}
  {\bibfnamefont {R.~W.}\ \bibnamefont {Morris}}, \bibinfo {author}
  {\bibfnamefont {A.}~\bibnamefont {Nethwewala}}, \bibinfo {author}
  {\bibfnamefont {S.}~\bibnamefont {Neyens}}, \bibinfo {author} {\bibfnamefont
  {R.}~\bibnamefont {Otten}}, \bibinfo {author} {\bibfnamefont {L.~P.}\
  \bibnamefont {Osuna~Ibarra}}, \bibinfo {author} {\bibfnamefont
  {B.}~\bibnamefont {Patra}}, \bibinfo {author} {\bibfnamefont
  {R.}~\bibnamefont {Pillarisetty}}, \bibinfo {author} {\bibfnamefont
  {S.}~\bibnamefont {Premaratne}}, \bibinfo {author} {\bibfnamefont
  {M.}~\bibnamefont {Ramsey}}, \bibinfo {author} {\bibfnamefont
  {A.}~\bibnamefont {Risinger}}, \bibinfo {author} {\bibfnamefont {J.~D.}\
  \bibnamefont {Rooney}}, \bibinfo {author} {\bibfnamefont {R.}~\bibnamefont
  {Savytskyy}}, \bibinfo {author} {\bibfnamefont {T.~F.}\ \bibnamefont
  {Watson}}, \bibinfo {author} {\bibfnamefont {O.~K.}\ \bibnamefont {Zietz}},
  \bibinfo {author} {\bibfnamefont {A.~Y.}\ \bibnamefont {Matsuura}}, \bibinfo
  {author} {\bibfnamefont {S.}~\bibnamefont {Pellerano}}, \bibinfo {author}
  {\bibfnamefont {N.~C.}\ \bibnamefont {Bishop}}, \bibinfo {author}
  {\bibfnamefont {J.}~\bibnamefont {Roberts}},\ and\ \bibinfo {author}
  {\bibfnamefont {J.~S.}\ \bibnamefont {Clarke}},\ }\bibfield  {title}
  {\bibinfo {title} {Operating two exchange-only qubits in parallel},\ }\href
  {https://doi.org/10.1038/s41586-025-09767-5} {\bibfield  {journal} {\bibinfo
  {journal} {Nature}\ }\textbf {\bibinfo {volume} {647}},\ \bibinfo {pages}
  {870} (\bibinfo {year} {2025})}\BibitemShut {NoStop}%
\bibitem [{\citenamefont {Broz}\ \emph
  {et~al.}(2026{\natexlab{a}})\citenamefont {Broz}, \citenamefont {Hoke},
  \citenamefont {Acuna},\ and\ \citenamefont {Petta}}]{Broz2025}%
  \BibitemOpen
  \bibfield  {author} {\bibinfo {author} {\bibfnamefont {J.~D.}\ \bibnamefont
  {Broz}}, \bibinfo {author} {\bibfnamefont {J.~C.}\ \bibnamefont {Hoke}},
  \bibinfo {author} {\bibfnamefont {E.}~\bibnamefont {Acuna}},\ and\ \bibinfo
  {author} {\bibfnamefont {J.~R.}\ \bibnamefont {Petta}},\ }\bibfield  {title}
  {\bibinfo {title} {Demonstration of an always-on exchange-only spin qubit},\
  }\href {https://doi.org/10.1038/s41467-026-70943-w} {\bibfield  {journal}
  {\bibinfo  {journal} {Nat. Commun.}\ }\textbf {\bibinfo {volume} {17}},\
  \bibinfo {pages} {4794} (\bibinfo {year} {2026}{\natexlab{a}})}\BibitemShut
  {NoStop}%
\bibitem [{\citenamefont {Broz}\ \emph
  {et~al.}(2026{\natexlab{b}})\citenamefont {Broz}, \citenamefont {Hoke},
  \citenamefont {Acuna},\ and\ \citenamefont {Petta}}]{Broz2026}%
  \BibitemOpen
  \bibfield  {author} {\bibinfo {author} {\bibfnamefont {J.~D.}\ \bibnamefont
  {Broz}}, \bibinfo {author} {\bibfnamefont {J.~C.}\ \bibnamefont {Hoke}},
  \bibinfo {author} {\bibfnamefont {E.}~\bibnamefont {Acuna}},\ and\ \bibinfo
  {author} {\bibfnamefont {J.~R.}\ \bibnamefont {Petta}},\ }\bibfield  {title}
  {\bibinfo {title} {Leakage-protected idle operation of a triangular
  exchange-only spin qubit},\ }\href
  {https://link.aps.org/doi/10.1103/sljc-cmdw} {\bibfield  {journal} {\bibinfo
  {journal} {PRX Quantum}\ }\textbf {\bibinfo {volume} {7}},\ \bibinfo {pages}
  {020324} (\bibinfo {year} {2026}{\natexlab{b}})}\BibitemShut {NoStop}%
\bibitem [{\citenamefont {Loss}\ and\ \citenamefont
  {DiVincenzo}(1998)}]{loss_quantum_1998}%
  \BibitemOpen
  \bibfield  {author} {\bibinfo {author} {\bibfnamefont {D.}~\bibnamefont
  {Loss}}\ and\ \bibinfo {author} {\bibfnamefont {D.~P.}\ \bibnamefont
  {DiVincenzo}},\ }\bibfield  {title} {\bibinfo {title} {Quantum computation
  with quantum dots},\ }\href {https://doi.org/10.1103/PhysRevA.57.120}
  {\bibfield  {journal} {\bibinfo  {journal} {Phys. Rev. A}\ }\textbf {\bibinfo
  {volume} {57}},\ \bibinfo {pages} {120} (\bibinfo {year} {1998})}\BibitemShut
  {NoStop}%
\bibitem [{\citenamefont {Bartee}\ \emph {et~al.}(2025)\citenamefont {Bartee},
  \citenamefont {Gilbert}, \citenamefont {Zuo}, \citenamefont {Das},
  \citenamefont {Tanttu}, \citenamefont {Yang}, \citenamefont
  {Dumoulin~Stuyck}, \citenamefont {Pauka}, \citenamefont {Su}, \citenamefont
  {Lim}, \citenamefont {Serrano}, \citenamefont {Escott}, \citenamefont
  {Hudson}, \citenamefont {Itoh}, \citenamefont {Laucht}, \citenamefont
  {Dzurak},\ and\ \citenamefont {Reilly}}]{bartee2025spinqubit}%
  \BibitemOpen
  \bibfield  {author} {\bibinfo {author} {\bibfnamefont {S.~K.}\ \bibnamefont
  {Bartee}}, \bibinfo {author} {\bibfnamefont {W.}~\bibnamefont {Gilbert}},
  \bibinfo {author} {\bibfnamefont {K.}~\bibnamefont {Zuo}}, \bibinfo {author}
  {\bibfnamefont {K.}~\bibnamefont {Das}}, \bibinfo {author} {\bibfnamefont
  {T.}~\bibnamefont {Tanttu}}, \bibinfo {author} {\bibfnamefont {C.~H.}\
  \bibnamefont {Yang}}, \bibinfo {author} {\bibfnamefont {N.}~\bibnamefont
  {Dumoulin~Stuyck}}, \bibinfo {author} {\bibfnamefont {S.~J.}\ \bibnamefont
  {Pauka}}, \bibinfo {author} {\bibfnamefont {R.~Y.}\ \bibnamefont {Su}},
  \bibinfo {author} {\bibfnamefont {W.~H.}\ \bibnamefont {Lim}}, \bibinfo
  {author} {\bibfnamefont {S.}~\bibnamefont {Serrano}}, \bibinfo {author}
  {\bibfnamefont {C.~C.}\ \bibnamefont {Escott}}, \bibinfo {author}
  {\bibfnamefont {F.~E.}\ \bibnamefont {Hudson}}, \bibinfo {author}
  {\bibfnamefont {K.~M.}\ \bibnamefont {Itoh}}, \bibinfo {author}
  {\bibfnamefont {A.}~\bibnamefont {Laucht}}, \bibinfo {author} {\bibfnamefont
  {A.~S.}\ \bibnamefont {Dzurak}},\ and\ \bibinfo {author} {\bibfnamefont
  {D.~J.}\ \bibnamefont {Reilly}},\ }\bibfield  {title} {\bibinfo {title}
  {Spin-qubit control with a milli-kelvin {CMOS} chip},\ }\href
  {https://doi.org/10.1038/s41586-025-09157-x} {\bibfield  {journal} {\bibinfo
  {journal} {Nature}\ }\textbf {\bibinfo {volume} {643}},\ \bibinfo {pages}
  {382} (\bibinfo {year} {2025})}\BibitemShut {NoStop}%
\bibitem [{\citenamefont {Undseth}\ \emph {et~al.}(2023)\citenamefont
  {Undseth}, \citenamefont {Xue}, \citenamefont {Mehmandoost}, \citenamefont
  {Rimbach-Russ}, \citenamefont {Eendebak}, \citenamefont {Samkharadze},
  \citenamefont {Sammak}, \citenamefont {Dobrovitski}, \citenamefont
  {Scappucci},\ and\ \citenamefont {Vandersypen}}]{Undseth2023}%
  \BibitemOpen
  \bibfield  {author} {\bibinfo {author} {\bibfnamefont {B.}~\bibnamefont
  {Undseth}}, \bibinfo {author} {\bibfnamefont {X.}~\bibnamefont {Xue}},
  \bibinfo {author} {\bibfnamefont {M.}~\bibnamefont {Mehmandoost}}, \bibinfo
  {author} {\bibfnamefont {M.}~\bibnamefont {Rimbach-Russ}}, \bibinfo {author}
  {\bibfnamefont {P.~T.}\ \bibnamefont {Eendebak}}, \bibinfo {author}
  {\bibfnamefont {N.}~\bibnamefont {Samkharadze}}, \bibinfo {author}
  {\bibfnamefont {A.}~\bibnamefont {Sammak}}, \bibinfo {author} {\bibfnamefont
  {V.~V.}\ \bibnamefont {Dobrovitski}}, \bibinfo {author} {\bibfnamefont
  {G.}~\bibnamefont {Scappucci}},\ and\ \bibinfo {author} {\bibfnamefont
  {L.~M.}\ \bibnamefont {Vandersypen}},\ }\bibfield  {title} {\bibinfo {title}
  {Nonlinear response and crosstalk of electrically driven silicon spin
  qubits},\ }\href {https://doi.org/10.1103/PhysRevApplied.19.044078}
  {\bibfield  {journal} {\bibinfo  {journal} {Phys. Rev. Appl.}\ }\textbf
  {\bibinfo {volume} {19}},\ \bibinfo {pages} {044078} (\bibinfo {year}
  {2023})}\BibitemShut {NoStop}%
\bibitem [{\citenamefont {Lawrie}\ \emph {et~al.}(2023)\citenamefont {Lawrie},
  \citenamefont {Rimbach-Russ}, \citenamefont {van Riggelen}, \citenamefont
  {Hendrickx}, \citenamefont {de~Snoo}, \citenamefont {Sammak}, \citenamefont
  {Scappucci}, \citenamefont {Helsen},\ and\ \citenamefont
  {Veldhorst}}]{Lawrie2023}%
  \BibitemOpen
  \bibfield  {author} {\bibinfo {author} {\bibfnamefont {W.~I.~L.}\
  \bibnamefont {Lawrie}}, \bibinfo {author} {\bibfnamefont {M.}~\bibnamefont
  {Rimbach-Russ}}, \bibinfo {author} {\bibfnamefont {F.}~\bibnamefont {van
  Riggelen}}, \bibinfo {author} {\bibfnamefont {N.~W.}\ \bibnamefont
  {Hendrickx}}, \bibinfo {author} {\bibfnamefont {S.~L.}\ \bibnamefont
  {de~Snoo}}, \bibinfo {author} {\bibfnamefont {A.}~\bibnamefont {Sammak}},
  \bibinfo {author} {\bibfnamefont {G.}~\bibnamefont {Scappucci}}, \bibinfo
  {author} {\bibfnamefont {J.}~\bibnamefont {Helsen}},\ and\ \bibinfo {author}
  {\bibfnamefont {M.}~\bibnamefont {Veldhorst}},\ }\bibfield  {title} {\bibinfo
  {title} {Simultaneous single-qubit driving of semiconductor spin qubits at
  the fault-tolerant threshold},\ }\href
  {http://dx.doi.org/10.1038/s41467-023-39334-3} {\bibfield  {journal}
  {\bibinfo  {journal} {Nat. Commun.}\ }\textbf {\bibinfo {volume} {14}},\
  \bibinfo {pages} {3617} (\bibinfo {year} {2023})}\BibitemShut {NoStop}%
\bibitem [{\citenamefont {DiVincenzo}\ \emph {et~al.}(2000)\citenamefont
  {DiVincenzo}, \citenamefont {Bacon}, \citenamefont {Kempe}, \citenamefont
  {Burkard},\ and\ \citenamefont {Whaley}}]{divincenzo_universal_2000}%
  \BibitemOpen
  \bibfield  {author} {\bibinfo {author} {\bibfnamefont {D.}~\bibnamefont
  {DiVincenzo}}, \bibinfo {author} {\bibfnamefont {D.}~\bibnamefont {Bacon}},
  \bibinfo {author} {\bibfnamefont {J.}~\bibnamefont {Kempe}}, \bibinfo
  {author} {\bibfnamefont {G.}~\bibnamefont {Burkard}},\ and\ \bibinfo {author}
  {\bibfnamefont {K.}~\bibnamefont {Whaley}},\ }\bibfield  {title} {\bibinfo
  {title} {Universal quantum computation with the exchange interaction},\
  }\href {https://www.nature.com/articles/35042541} {\bibfield  {journal}
  {\bibinfo  {journal} {Nature}\ }\textbf {\bibinfo {volume} {408}},\ \bibinfo
  {pages} {339} (\bibinfo {year} {2000})}\BibitemShut {NoStop}%
\bibitem [{\citenamefont {Fong}\ and\ \citenamefont
  {Wandzura}(2011)}]{fong_universal_2011}%
  \BibitemOpen
  \bibfield  {author} {\bibinfo {author} {\bibfnamefont {B.~H.}\ \bibnamefont
  {Fong}}\ and\ \bibinfo {author} {\bibfnamefont {S.~M.}\ \bibnamefont
  {Wandzura}},\ }\bibfield  {title} {\bibinfo {title} {Universal quantum
  computation and leakage reduction in the 3-qubit decoherence free
  subsystem},\ }\href {https://dl.acm.org/doi/abs/10.5555/2230956.2230965}
  {\bibfield  {journal} {\bibinfo  {journal} {Quantum Info. Comput.}\ }\textbf
  {\bibinfo {volume} {11}},\ \bibinfo {pages} {1003} (\bibinfo {year}
  {2011})}\BibitemShut {NoStop}%
\bibitem [{\citenamefont {Sala}\ and\ \citenamefont
  {Danon}(2017)}]{sala_exchange-only_2017}%
  \BibitemOpen
  \bibfield  {author} {\bibinfo {author} {\bibfnamefont {A.}~\bibnamefont
  {Sala}}\ and\ \bibinfo {author} {\bibfnamefont {J.}~\bibnamefont {Danon}},\
  }\bibfield  {title} {\bibinfo {title} {Exchange-only singlet-only spin
  qubit},\ }\href {https://doi.org/10.1103/PhysRevB.95.241303} {\bibfield
  {journal} {\bibinfo  {journal} {Phys. Rev. B}\ }\textbf {\bibinfo {volume}
  {95}},\ \bibinfo {pages} {241303(R)} (\bibinfo {year} {2017})}\BibitemShut
  {NoStop}%
\bibitem [{\citenamefont {Bacon}\ \emph {et~al.}(1999)\citenamefont {Bacon},
  \citenamefont {Lidar},\ and\ \citenamefont {Whaley}}]{bacon_robustness_1999}%
  \BibitemOpen
  \bibfield  {author} {\bibinfo {author} {\bibfnamefont {D.}~\bibnamefont
  {Bacon}}, \bibinfo {author} {\bibfnamefont {D.~A.}\ \bibnamefont {Lidar}},\
  and\ \bibinfo {author} {\bibfnamefont {K.~B.}\ \bibnamefont {Whaley}},\
  }\bibfield  {title} {\bibinfo {title} {Robustness of decoherence-free
  subspaces for quantum computation},\ }\href
  {https://doi.org/10.1103/PhysRevA.60.1944} {\bibfield  {journal} {\bibinfo
  {journal} {Phys. Rev. A}\ }\textbf {\bibinfo {volume} {60}},\ \bibinfo
  {pages} {1944} (\bibinfo {year} {1999})}\BibitemShut {NoStop}%
\bibitem [{\citenamefont {Bacon}\ \emph {et~al.}(2000)\citenamefont {Bacon},
  \citenamefont {Kempe}, \citenamefont {Lidar},\ and\ \citenamefont
  {Whaley}}]{bacon_universal_2000}%
  \BibitemOpen
  \bibfield  {author} {\bibinfo {author} {\bibfnamefont {D.}~\bibnamefont
  {Bacon}}, \bibinfo {author} {\bibfnamefont {J.}~\bibnamefont {Kempe}},
  \bibinfo {author} {\bibfnamefont {D.~A.}\ \bibnamefont {Lidar}},\ and\
  \bibinfo {author} {\bibfnamefont {K.~B.}\ \bibnamefont {Whaley}},\ }\bibfield
   {title} {\bibinfo {title} {Universal {Fault}-{Tolerant} {Computation} on
  {Decoherence}-{Free} {Subspaces}},\ }\href
  {https://doi.org/10.1103/PhysRevLett.85.1758} {\bibfield  {journal} {\bibinfo
   {journal} {Phys. Rev. Lett.}\ }\textbf {\bibinfo {volume} {85}},\ \bibinfo
  {pages} {1758} (\bibinfo {year} {2000})}\BibitemShut {NoStop}%
\bibitem [{\citenamefont {Bacon}\ \emph {et~al.}(2001)\citenamefont {Bacon},
  \citenamefont {Kempe}, \citenamefont {DiVincenzo}, \citenamefont {Lidar},\
  and\ \citenamefont {Whaley}}]{bacon_encoded_2001}%
  \BibitemOpen
  \bibfield  {author} {\bibinfo {author} {\bibfnamefont {D.}~\bibnamefont
  {Bacon}}, \bibinfo {author} {\bibfnamefont {J.}~\bibnamefont {Kempe}},
  \bibinfo {author} {\bibfnamefont {D.~P.}\ \bibnamefont {DiVincenzo}},
  \bibinfo {author} {\bibfnamefont {D.~A.}\ \bibnamefont {Lidar}},\ and\
  \bibinfo {author} {\bibfnamefont {K.~B.}\ \bibnamefont {Whaley}},\ }\bibfield
   {title} {\bibinfo {title} {Encoded universality in physical implementations
  of a quantum computer},\ }in\ \href {https://arxiv.org/abs/quant-ph/0102140}
  {\emph {\bibinfo {booktitle} {Proceedings of the 1st International Conference
  on Experimental Implementations of Quantum Computation}}},\ \bibinfo {editor}
  {edited by\ \bibinfo {editor} {\bibfnamefont {R.}~\bibnamefont {Clark}}}\
  (\bibinfo  {publisher} {Rinton},\ \bibinfo {address} {Princeton, NJ},\
  \bibinfo {year} {2001})\ p.\ \bibinfo {pages} {257}\BibitemShut {NoStop}%
\bibitem [{\citenamefont {Kempe}\ \emph
  {et~al.}(2001{\natexlab{a}})\citenamefont {Kempe}, \citenamefont {Bacon},
  \citenamefont {DiVincenzo},\ and\ \citenamefont
  {Whaley}}]{kempe_encoded_2001}%
  \BibitemOpen
  \bibfield  {author} {\bibinfo {author} {\bibfnamefont {J.}~\bibnamefont
  {Kempe}}, \bibinfo {author} {\bibfnamefont {D.}~\bibnamefont {Bacon}},
  \bibinfo {author} {\bibfnamefont {D.~P.}\ \bibnamefont {DiVincenzo}},\ and\
  \bibinfo {author} {\bibfnamefont {K.~B.}\ \bibnamefont {Whaley}},\ }\bibfield
   {title} {\bibinfo {title} {Encoded universality from a single physical
  interaction},\ }\href {https://dl.acm.org/doi/abs/10.5555/2016994.2017000}
  {\bibfield  {journal} {\bibinfo  {journal} {Quantum Info. Comput.}\ }\textbf
  {\bibinfo {volume} {1}},\ \bibinfo {pages} {33} (\bibinfo {year}
  {2001}{\natexlab{a}})}\BibitemShut {NoStop}%
\bibitem [{\citenamefont {Kempe}\ \emph
  {et~al.}(2001{\natexlab{b}})\citenamefont {Kempe}, \citenamefont {Bacon},
  \citenamefont {Lidar},\ and\ \citenamefont {Whaley}}]{kempe_theory_2001}%
  \BibitemOpen
  \bibfield  {author} {\bibinfo {author} {\bibfnamefont {J.}~\bibnamefont
  {Kempe}}, \bibinfo {author} {\bibfnamefont {D.}~\bibnamefont {Bacon}},
  \bibinfo {author} {\bibfnamefont {D.~A.}\ \bibnamefont {Lidar}},\ and\
  \bibinfo {author} {\bibfnamefont {K.~B.}\ \bibnamefont {Whaley}},\ }\bibfield
   {title} {\bibinfo {title} {Theory of decoherence-free fault-tolerant
  universal quantum computation},\ }\href
  {https://doi.org/10.1103/PhysRevA.63.042307} {\bibfield  {journal} {\bibinfo
  {journal} {Phys. Rev. A}\ }\textbf {\bibinfo {volume} {63}},\ \bibinfo
  {pages} {042307} (\bibinfo {year} {2001}{\natexlab{b}})}\BibitemShut
  {NoStop}%
\bibitem [{\citenamefont {Bacon}(2001)}]{bacon_decoherence_2003}%
  \BibitemOpen
  \bibfield  {author} {\bibinfo {author} {\bibfnamefont {D.}~\bibnamefont
  {Bacon}},\ }\emph {\bibinfo {title} {Decoherence, {Control}, and {Symmetry}
  in {Quantum} {Computers}}},\ \href {http://arxiv.org/abs/quant-ph/0305025}
  {\bibinfo {type} {{Ph.D.} thesis}},\ \bibinfo  {school} {University of
  California, Berkeley} (\bibinfo {year} {2001})\BibitemShut {NoStop}%
\bibitem [{\citenamefont {Sala}\ \emph {et~al.}(2020)\citenamefont {Sala},
  \citenamefont {Qvist},\ and\ \citenamefont {Danon}}]{sala_highly_2020}%
  \BibitemOpen
  \bibfield  {author} {\bibinfo {author} {\bibfnamefont {A.}~\bibnamefont
  {Sala}}, \bibinfo {author} {\bibfnamefont {J.~H.}\ \bibnamefont {Qvist}},\
  and\ \bibinfo {author} {\bibfnamefont {J.}~\bibnamefont {Danon}},\ }\bibfield
   {title} {\bibinfo {title} {Highly tunable exchange-only singlet-only qubit
  in a {GaAs} triple quantum dot},\ }\href
  {https://doi.org/10.1103/PhysRevResearch.2.012062} {\bibfield  {journal}
  {\bibinfo  {journal} {Phys. Rev. Research}\ }\textbf {\bibinfo {volume}
  {2}},\ \bibinfo {pages} {012062(R)} (\bibinfo {year} {2020})}\BibitemShut
  {NoStop}%
\bibitem [{\citenamefont {Weinstein}\ and\ \citenamefont
  {Hellberg}(2005)}]{weinstein_energetic_2005}%
  \BibitemOpen
  \bibfield  {author} {\bibinfo {author} {\bibfnamefont {Y.~S.}\ \bibnamefont
  {Weinstein}}\ and\ \bibinfo {author} {\bibfnamefont {C.~S.}\ \bibnamefont
  {Hellberg}},\ }\bibfield  {title} {\bibinfo {title} {Energetic suppression of
  decoherence in exchange-only quantum computation},\ }\href
  {https://doi.org/10.1103/PhysRevA.72.022319} {\bibfield  {journal} {\bibinfo
  {journal} {Phys. Rev. A}\ }\textbf {\bibinfo {volume} {72}},\ \bibinfo
  {pages} {022319} (\bibinfo {year} {2005})}\BibitemShut {NoStop}%
\bibitem [{\citenamefont {Foulk}\ \emph {et~al.}(2025)\citenamefont {Foulk},
  \citenamefont {Hoffman}, \citenamefont {Laubscher},\ and\ \citenamefont
  {Das~Sarma}}]{Foulk2025}%
  \BibitemOpen
  \bibfield  {author} {\bibinfo {author} {\bibfnamefont {N.~L.}\ \bibnamefont
  {Foulk}}, \bibinfo {author} {\bibfnamefont {S.}~\bibnamefont {Hoffman}},
  \bibinfo {author} {\bibfnamefont {K.}~\bibnamefont {Laubscher}},\ and\
  \bibinfo {author} {\bibfnamefont {S.}~\bibnamefont {Das~Sarma}},\ }\bibfield
  {title} {\bibinfo {title} {Singlet-only always-on gapless exchange qubits
  with baseband control},\ }\href {https://link.aps.org/doi/10.1103/j1hn-k5m8}
  {\bibfield  {journal} {\bibinfo  {journal} {Phys. Rev. Lett.}\ }\textbf
  {\bibinfo {volume} {135}},\ \bibinfo {pages} {106202} (\bibinfo {year}
  {2025})}\BibitemShut {NoStop}%
\bibitem [{\citenamefont {Yang}\ \emph {et~al.}(2011)\citenamefont {Yang},
  \citenamefont {Wang},\ and\ \citenamefont {Das~Sarma}}]{yang2011generic}%
  \BibitemOpen
  \bibfield  {author} {\bibinfo {author} {\bibfnamefont {S.}~\bibnamefont
  {Yang}}, \bibinfo {author} {\bibfnamefont {X.}~\bibnamefont {Wang}},\ and\
  \bibinfo {author} {\bibfnamefont {S.}~\bibnamefont {Das~Sarma}},\ }\bibfield
  {title} {\bibinfo {title} {Generic {Hubbard} model description of
  semiconductor quantum-dot spin qubits},\ }\href
  {https://link.aps.org/doi/10.1103/PhysRevB.83.161301} {\bibfield  {journal}
  {\bibinfo  {journal} {Phys. Rev. B}\ }\textbf {\bibinfo {volume} {83}},\
  \bibinfo {pages} {161301} (\bibinfo {year} {2011})}\BibitemShut {NoStop}%
\bibitem [{\citenamefont {Wang}\ \emph {et~al.}(2011)\citenamefont {Wang},
  \citenamefont {Yang},\ and\ \citenamefont {Das~Sarma}}]{yang2011quantum}%
  \BibitemOpen
  \bibfield  {author} {\bibinfo {author} {\bibfnamefont {X.}~\bibnamefont
  {Wang}}, \bibinfo {author} {\bibfnamefont {S.}~\bibnamefont {Yang}},\ and\
  \bibinfo {author} {\bibfnamefont {S.}~\bibnamefont {Das~Sarma}},\ }\bibfield
  {title} {\bibinfo {title} {Quantum theory of the charge-stability diagram of
  semiconductor double-quantum-dot systems},\ }\href
  {https://doi.org/10.1103/PhysRevB.84.115301} {\bibfield  {journal} {\bibinfo
  {journal} {Phys. Rev. B}\ }\textbf {\bibinfo {volume} {84}},\ \bibinfo
  {pages} {115301} (\bibinfo {year} {2011})}\BibitemShut {NoStop}%
\bibitem [{\citenamefont {Das~Sarma}\ \emph {et~al.}(2011)\citenamefont
  {Das~Sarma}, \citenamefont {Wang},\ and\ \citenamefont
  {Yang}}]{Dassarma2011}%
  \BibitemOpen
  \bibfield  {author} {\bibinfo {author} {\bibfnamefont {S.}~\bibnamefont
  {Das~Sarma}}, \bibinfo {author} {\bibfnamefont {X.}~\bibnamefont {Wang}},\
  and\ \bibinfo {author} {\bibfnamefont {S.}~\bibnamefont {Yang}},\ }\bibfield
  {title} {\bibinfo {title} {Hubbard model description of silicon spin qubits:
  {Charge} stability diagram and tunnel coupling in {Si} double quantum dots},\
  }\href {https://doi.org/10.1103/PhysRevB.83.235314} {\bibfield  {journal}
  {\bibinfo  {journal} {Phys. Rev. B}\ }\textbf {\bibinfo {volume} {83}},\
  \bibinfo {pages} {235314} (\bibinfo {year} {2011})}\BibitemShut {NoStop}%
\bibitem [{\citenamefont {Carr}\ and\ \citenamefont
  {Purcell}(1954)}]{Carr1954Effects}%
  \BibitemOpen
  \bibfield  {author} {\bibinfo {author} {\bibfnamefont {H.~Y.}\ \bibnamefont
  {Carr}}\ and\ \bibinfo {author} {\bibfnamefont {E.~M.}\ \bibnamefont
  {Purcell}},\ }\bibfield  {title} {\bibinfo {title} {Effects of diffusion on
  free precession in nuclear magnetic resonance experiments},\ }\href
  {https://doi.org/10.1103/PhysRev.94.630} {\bibfield  {journal} {\bibinfo
  {journal} {Physical Review}\ }\textbf {\bibinfo {volume} {94}},\ \bibinfo
  {pages} {630} (\bibinfo {year} {1954})}\BibitemShut {NoStop}%
\bibitem [{\citenamefont {Meiboom}\ and\ \citenamefont
  {Gill}(1958)}]{Meiboom1958Modified}%
  \BibitemOpen
  \bibfield  {author} {\bibinfo {author} {\bibfnamefont {S.}~\bibnamefont
  {Meiboom}}\ and\ \bibinfo {author} {\bibfnamefont {D.}~\bibnamefont {Gill}},\
  }\bibfield  {title} {\bibinfo {title} {Modified spin‐echo method for
  measuring nuclear relaxation times},\ }\href
  {https://doi.org/10.1063/1.1716296} {\bibfield  {journal} {\bibinfo
  {journal} {Review of Scientific Instruments}\ }\textbf {\bibinfo {volume}
  {29}},\ \bibinfo {pages} {688} (\bibinfo {year} {1958})}\BibitemShut
  {NoStop}%
\bibitem [{\citenamefont {Shim}\ and\ \citenamefont
  {Tahan}(2016)}]{shim_charge-noise-insensitive_2016}%
  \BibitemOpen
  \bibfield  {author} {\bibinfo {author} {\bibfnamefont {Y.-P.}\ \bibnamefont
  {Shim}}\ and\ \bibinfo {author} {\bibfnamefont {C.}~\bibnamefont {Tahan}},\
  }\bibfield  {title} {\bibinfo {title} {Charge-noise-insensitive gate
  operations for always-on, exchange-only qubits},\ }\href
  {https://link.aps.org/doi/10.1103/PhysRevB.93.121410} {\bibfield  {journal}
  {\bibinfo  {journal} {Phys. Rev. B}\ }\textbf {\bibinfo {volume} {93}},\
  \bibinfo {pages} {121410(R)} (\bibinfo {year} {2016})}\BibitemShut {NoStop}%
\bibitem [{\citenamefont {Reed}\ \emph {et~al.}(2016)\citenamefont {Reed},
  \citenamefont {Maune}, \citenamefont {Andrews}, \citenamefont {Borselli},
  \citenamefont {Eng}, \citenamefont {Jura}, \citenamefont {Kiselev},
  \citenamefont {Ladd}, \citenamefont {Merkel}, \citenamefont {Milosavljevic},
  \citenamefont {Pritchett}, \citenamefont {Rakher}, \citenamefont {Ross},
  \citenamefont {Schmitz}, \citenamefont {Smith}, \citenamefont {Wright},
  \citenamefont {Gyure},\ and\ \citenamefont {Hunter}}]{reed_reduced_2016}%
  \BibitemOpen
  \bibfield  {author} {\bibinfo {author} {\bibfnamefont {M.~D.}\ \bibnamefont
  {Reed}}, \bibinfo {author} {\bibfnamefont {B.~M.}\ \bibnamefont {Maune}},
  \bibinfo {author} {\bibfnamefont {R.~W.}\ \bibnamefont {Andrews}}, \bibinfo
  {author} {\bibfnamefont {M.~G.}\ \bibnamefont {Borselli}}, \bibinfo {author}
  {\bibfnamefont {K.}~\bibnamefont {Eng}}, \bibinfo {author} {\bibfnamefont
  {M.~P.}\ \bibnamefont {Jura}}, \bibinfo {author} {\bibfnamefont {A.~A.}\
  \bibnamefont {Kiselev}}, \bibinfo {author} {\bibfnamefont {T.~D.}\
  \bibnamefont {Ladd}}, \bibinfo {author} {\bibfnamefont {S.~T.}\ \bibnamefont
  {Merkel}}, \bibinfo {author} {\bibfnamefont {I.}~\bibnamefont
  {Milosavljevic}}, \bibinfo {author} {\bibfnamefont {E.~J.}\ \bibnamefont
  {Pritchett}}, \bibinfo {author} {\bibfnamefont {M.~T.}\ \bibnamefont
  {Rakher}}, \bibinfo {author} {\bibfnamefont {R.~S.}\ \bibnamefont {Ross}},
  \bibinfo {author} {\bibfnamefont {A.~E.}\ \bibnamefont {Schmitz}}, \bibinfo
  {author} {\bibfnamefont {A.}~\bibnamefont {Smith}}, \bibinfo {author}
  {\bibfnamefont {J.~A.}\ \bibnamefont {Wright}}, \bibinfo {author}
  {\bibfnamefont {M.~F.}\ \bibnamefont {Gyure}},\ and\ \bibinfo {author}
  {\bibfnamefont {A.~T.}\ \bibnamefont {Hunter}},\ }\bibfield  {title}
  {\bibinfo {title} {Reduced sensitivity to charge noise in semiconductor spin
  qubits via symmetric operation},\ }\href
  {https://doi.org/10.1103/PhysRevLett.116.110402} {\bibfield  {journal}
  {\bibinfo  {journal} {Phys. Rev. Lett.}\ }\textbf {\bibinfo {volume} {116}},\
  \bibinfo {pages} {110402} (\bibinfo {year} {2016})}\BibitemShut {NoStop}%
\bibitem [{\citenamefont {Freeman}\ \emph {et~al.}(2016)\citenamefont
  {Freeman}, \citenamefont {Schoenfield},\ and\ \citenamefont
  {Jiang}}]{Freeman2016}%
  \BibitemOpen
  \bibfield  {author} {\bibinfo {author} {\bibfnamefont {B.~M.}\ \bibnamefont
  {Freeman}}, \bibinfo {author} {\bibfnamefont {J.~S.}\ \bibnamefont
  {Schoenfield}},\ and\ \bibinfo {author} {\bibfnamefont {H.}~\bibnamefont
  {Jiang}},\ }\bibfield  {title} {\bibinfo {title} {Comparison of low frequency
  charge noise in identically patterned {Si/SiO2} and {Si/SiGe} quantum dots},\
  }\href {https://doi.org/10.1063/1.4954700} {\bibfield  {journal} {\bibinfo
  {journal} {Appl. Phys. Lett.}\ }\textbf {\bibinfo {volume} {108}},\ \bibinfo
  {pages} {253108} (\bibinfo {year} {2016})}\BibitemShut {NoStop}%
\bibitem [{\citenamefont {Connors}\ \emph {et~al.}(2022)\citenamefont
  {Connors}, \citenamefont {Nelson}, \citenamefont {Edge},\ and\ \citenamefont
  {Nichol}}]{Connors2022}%
  \BibitemOpen
  \bibfield  {author} {\bibinfo {author} {\bibfnamefont {E.~J.}\ \bibnamefont
  {Connors}}, \bibinfo {author} {\bibfnamefont {J.}~\bibnamefont {Nelson}},
  \bibinfo {author} {\bibfnamefont {L.~F.}\ \bibnamefont {Edge}},\ and\
  \bibinfo {author} {\bibfnamefont {J.~M.}\ \bibnamefont {Nichol}},\ }\bibfield
   {title} {\bibinfo {title} {Charge-noise spectroscopy of {Si/SiGe} quantum
  dots via dynamically-decoupled exchange oscillations},\ }\href
  {http://dx.doi.org/10.1038/s41467-022-28519-x} {\bibfield  {journal}
  {\bibinfo  {journal} {Nat. Commun.}\ }\textbf {\bibinfo {volume} {13}},\
  \bibinfo {pages} {940} (\bibinfo {year} {2022})}\BibitemShut {NoStop}%
\bibitem [{\citenamefont {Steinacker}\ \emph {et~al.}(2025)\citenamefont
  {Steinacker}, \citenamefont {Dumoulin~Stuyck}, \citenamefont {Lim},
  \citenamefont {Tanttu}, \citenamefont {Feng}, \citenamefont {Serrano},
  \citenamefont {Nickl}, \citenamefont {Candido}, \citenamefont {Cifuentes},
  \citenamefont {Vahapoglu}, \citenamefont {Bartee}, \citenamefont {Hudson},
  \citenamefont {Chan}, \citenamefont {Kubicek}, \citenamefont {Jussot},
  \citenamefont {Canvel}, \citenamefont {Beyne}, \citenamefont {Shimura},
  \citenamefont {Loo}, \citenamefont {Godfrin}, \citenamefont {Raes},
  \citenamefont {Baudot}, \citenamefont {Wan}, \citenamefont {Laucht},
  \citenamefont {Yang}, \citenamefont {Saraiva}, \citenamefont {Escott},
  \citenamefont {De~Greve},\ and\ \citenamefont {Dzurak}}]{Steinacker2025}%
  \BibitemOpen
  \bibfield  {author} {\bibinfo {author} {\bibfnamefont {P.}~\bibnamefont
  {Steinacker}}, \bibinfo {author} {\bibfnamefont {N.}~\bibnamefont
  {Dumoulin~Stuyck}}, \bibinfo {author} {\bibfnamefont {W.~H.}\ \bibnamefont
  {Lim}}, \bibinfo {author} {\bibfnamefont {T.}~\bibnamefont {Tanttu}},
  \bibinfo {author} {\bibfnamefont {M.}~\bibnamefont {Feng}}, \bibinfo {author}
  {\bibfnamefont {S.}~\bibnamefont {Serrano}}, \bibinfo {author} {\bibfnamefont
  {A.}~\bibnamefont {Nickl}}, \bibinfo {author} {\bibfnamefont
  {M.}~\bibnamefont {Candido}}, \bibinfo {author} {\bibfnamefont {J.~D.}\
  \bibnamefont {Cifuentes}}, \bibinfo {author} {\bibfnamefont {E.}~\bibnamefont
  {Vahapoglu}}, \bibinfo {author} {\bibfnamefont {S.~K.}\ \bibnamefont
  {Bartee}}, \bibinfo {author} {\bibfnamefont {F.~E.}\ \bibnamefont {Hudson}},
  \bibinfo {author} {\bibfnamefont {K.~W.}\ \bibnamefont {Chan}}, \bibinfo
  {author} {\bibfnamefont {S.}~\bibnamefont {Kubicek}}, \bibinfo {author}
  {\bibfnamefont {J.}~\bibnamefont {Jussot}}, \bibinfo {author} {\bibfnamefont
  {Y.}~\bibnamefont {Canvel}}, \bibinfo {author} {\bibfnamefont
  {S.}~\bibnamefont {Beyne}}, \bibinfo {author} {\bibfnamefont
  {Y.}~\bibnamefont {Shimura}}, \bibinfo {author} {\bibfnamefont
  {R.}~\bibnamefont {Loo}}, \bibinfo {author} {\bibfnamefont {C.}~\bibnamefont
  {Godfrin}}, \bibinfo {author} {\bibfnamefont {B.}~\bibnamefont {Raes}},
  \bibinfo {author} {\bibfnamefont {S.}~\bibnamefont {Baudot}}, \bibinfo
  {author} {\bibfnamefont {D.}~\bibnamefont {Wan}}, \bibinfo {author}
  {\bibfnamefont {A.}~\bibnamefont {Laucht}}, \bibinfo {author} {\bibfnamefont
  {C.~H.}\ \bibnamefont {Yang}}, \bibinfo {author} {\bibfnamefont
  {A.}~\bibnamefont {Saraiva}}, \bibinfo {author} {\bibfnamefont {C.~C.}\
  \bibnamefont {Escott}}, \bibinfo {author} {\bibfnamefont {K.}~\bibnamefont
  {De~Greve}},\ and\ \bibinfo {author} {\bibfnamefont {A.~S.}\ \bibnamefont
  {Dzurak}},\ }\bibfield  {title} {\bibinfo {title} {Industry-compatible
  silicon spin-qubit unit cells exceeding 99\% fidelity},\ }\href
  {http://dx.doi.org/10.1038/s41586-025-09531-9} {\bibfield  {journal}
  {\bibinfo  {journal} {Nature}\ }\textbf {\bibinfo {volume} {646}},\ \bibinfo
  {pages} {81} (\bibinfo {year} {2025})}\BibitemShut {NoStop}%
\bibitem [{\citenamefont {Elsayed}\ \emph {et~al.}(2024)\citenamefont
  {Elsayed}, \citenamefont {Shehata}, \citenamefont {Godfrin}, \citenamefont
  {Kubicek}, \citenamefont {Massar}, \citenamefont {Canvel}, \citenamefont
  {Jussot}, \citenamefont {Simion}, \citenamefont {Mongillo}, \citenamefont
  {Wan}, \citenamefont {Govoreanu}, \citenamefont {Radu}, \citenamefont {Li},
  \citenamefont {Van~Dorpe},\ and\ \citenamefont
  {De~Greve}}]{elsayed_low_2022}%
  \BibitemOpen
  \bibfield  {author} {\bibinfo {author} {\bibfnamefont {A.}~\bibnamefont
  {Elsayed}}, \bibinfo {author} {\bibfnamefont {M.~M.~K.}\ \bibnamefont
  {Shehata}}, \bibinfo {author} {\bibfnamefont {C.}~\bibnamefont {Godfrin}},
  \bibinfo {author} {\bibfnamefont {S.}~\bibnamefont {Kubicek}}, \bibinfo
  {author} {\bibfnamefont {S.}~\bibnamefont {Massar}}, \bibinfo {author}
  {\bibfnamefont {Y.}~\bibnamefont {Canvel}}, \bibinfo {author} {\bibfnamefont
  {J.}~\bibnamefont {Jussot}}, \bibinfo {author} {\bibfnamefont
  {G.}~\bibnamefont {Simion}}, \bibinfo {author} {\bibfnamefont
  {M.}~\bibnamefont {Mongillo}}, \bibinfo {author} {\bibfnamefont
  {D.}~\bibnamefont {Wan}}, \bibinfo {author} {\bibfnamefont {B.}~\bibnamefont
  {Govoreanu}}, \bibinfo {author} {\bibfnamefont {I.~P.}\ \bibnamefont {Radu}},
  \bibinfo {author} {\bibfnamefont {R.}~\bibnamefont {Li}}, \bibinfo {author}
  {\bibfnamefont {P.}~\bibnamefont {Van~Dorpe}},\ and\ \bibinfo {author}
  {\bibfnamefont {K.}~\bibnamefont {De~Greve}},\ }\bibfield  {title} {\bibinfo
  {title} {Low charge noise quantum dots with industrial {CMOS}
  manufacturing},\ }\href {http://dx.doi.org/10.1038/s41534-024-00864-3}
  {\bibfield  {journal} {\bibinfo  {journal} {npj Quantum Inf.}\ }\textbf
  {\bibinfo {volume} {10}},\ \bibinfo {pages} {70} (\bibinfo {year}
  {2024})}\BibitemShut {NoStop}%
\bibitem [{\citenamefont {Neyens}\ \emph {et~al.}(2024)\citenamefont {Neyens},
  \citenamefont {Zietz}, \citenamefont {Watson}, \citenamefont {Luthi},
  \citenamefont {Nethwewala}, \citenamefont {George}, \citenamefont {Henry},
  \citenamefont {Islam}, \citenamefont {Wagner}, \citenamefont {Borjans},
  \citenamefont {Connors}, \citenamefont {Corrigan}, \citenamefont {Curry},
  \citenamefont {Keith}, \citenamefont {Kotlyar}, \citenamefont {Lampert},
  \citenamefont {Mądzik}, \citenamefont {Millard}, \citenamefont {Mohiyaddin},
  \citenamefont {Pellerano}, \citenamefont {Pillarisetty}, \citenamefont
  {Ramsey}, \citenamefont {Savytskyy}, \citenamefont {Schaal}, \citenamefont
  {Zheng}, \citenamefont {Ziegler}, \citenamefont {Bishop}, \citenamefont
  {Bojarski}, \citenamefont {Roberts},\ and\ \citenamefont
  {Clarke}}]{neyens_probing_2024}%
  \BibitemOpen
  \bibfield  {author} {\bibinfo {author} {\bibfnamefont {S.}~\bibnamefont
  {Neyens}}, \bibinfo {author} {\bibfnamefont {O.}~\bibnamefont {Zietz}},
  \bibinfo {author} {\bibfnamefont {T.}~\bibnamefont {Watson}}, \bibinfo
  {author} {\bibfnamefont {F.}~\bibnamefont {Luthi}}, \bibinfo {author}
  {\bibfnamefont {A.}~\bibnamefont {Nethwewala}}, \bibinfo {author}
  {\bibfnamefont {H.}~\bibnamefont {George}}, \bibinfo {author} {\bibfnamefont
  {E.}~\bibnamefont {Henry}}, \bibinfo {author} {\bibfnamefont
  {M.}~\bibnamefont {Islam}}, \bibinfo {author} {\bibfnamefont
  {A.}~\bibnamefont {Wagner}}, \bibinfo {author} {\bibfnamefont
  {F.}~\bibnamefont {Borjans}}, \bibinfo {author} {\bibfnamefont
  {E.}~\bibnamefont {Connors}}, \bibinfo {author} {\bibfnamefont
  {J.}~\bibnamefont {Corrigan}}, \bibinfo {author} {\bibfnamefont
  {M.}~\bibnamefont {Curry}}, \bibinfo {author} {\bibfnamefont
  {D.}~\bibnamefont {Keith}}, \bibinfo {author} {\bibfnamefont
  {R.}~\bibnamefont {Kotlyar}}, \bibinfo {author} {\bibfnamefont
  {L.}~\bibnamefont {Lampert}}, \bibinfo {author} {\bibfnamefont
  {M.}~\bibnamefont {Mądzik}}, \bibinfo {author} {\bibfnamefont
  {K.}~\bibnamefont {Millard}}, \bibinfo {author} {\bibfnamefont
  {F.}~\bibnamefont {Mohiyaddin}}, \bibinfo {author} {\bibfnamefont
  {S.}~\bibnamefont {Pellerano}}, \bibinfo {author} {\bibfnamefont
  {R.}~\bibnamefont {Pillarisetty}}, \bibinfo {author} {\bibfnamefont
  {M.}~\bibnamefont {Ramsey}}, \bibinfo {author} {\bibfnamefont
  {R.}~\bibnamefont {Savytskyy}}, \bibinfo {author} {\bibfnamefont
  {S.}~\bibnamefont {Schaal}}, \bibinfo {author} {\bibfnamefont
  {G.}~\bibnamefont {Zheng}}, \bibinfo {author} {\bibfnamefont
  {J.}~\bibnamefont {Ziegler}}, \bibinfo {author} {\bibfnamefont
  {N.}~\bibnamefont {Bishop}}, \bibinfo {author} {\bibfnamefont
  {S.}~\bibnamefont {Bojarski}}, \bibinfo {author} {\bibfnamefont
  {J.}~\bibnamefont {Roberts}},\ and\ \bibinfo {author} {\bibfnamefont
  {J.}~\bibnamefont {Clarke}},\ }\bibfield  {title} {\bibinfo {title} {Probing
  single electrons across 300-mm spin qubit wafers},\ }\href
  {https://www.nature.com/articles/s41586-024-07275-6} {\bibfield  {journal}
  {\bibinfo  {journal} {Nature}\ }\textbf {\bibinfo {volume} {629}},\ \bibinfo
  {pages} {80} (\bibinfo {year} {2024})}\BibitemShut {NoStop}%
\bibitem [{\citenamefont {Koch}\ \emph {et~al.}(2025)\citenamefont {Koch},
  \citenamefont {Godfrin}, \citenamefont {Adam}, \citenamefont {Ferrero},
  \citenamefont {Schroller}, \citenamefont {Glaeser}, \citenamefont {Kubicek},
  \citenamefont {Li}, \citenamefont {Loo}, \citenamefont {Massar},
  \citenamefont {Simion}, \citenamefont {Wan}, \citenamefont {De~Greve},\ and\
  \citenamefont {Wernsdorfer}}]{Koch2025}%
  \BibitemOpen
  \bibfield  {author} {\bibinfo {author} {\bibfnamefont {T.}~\bibnamefont
  {Koch}}, \bibinfo {author} {\bibfnamefont {C.}~\bibnamefont {Godfrin}},
  \bibinfo {author} {\bibfnamefont {V.}~\bibnamefont {Adam}}, \bibinfo {author}
  {\bibfnamefont {J.}~\bibnamefont {Ferrero}}, \bibinfo {author} {\bibfnamefont
  {D.}~\bibnamefont {Schroller}}, \bibinfo {author} {\bibfnamefont
  {N.}~\bibnamefont {Glaeser}}, \bibinfo {author} {\bibfnamefont
  {S.}~\bibnamefont {Kubicek}}, \bibinfo {author} {\bibfnamefont
  {R.}~\bibnamefont {Li}}, \bibinfo {author} {\bibfnamefont {R.}~\bibnamefont
  {Loo}}, \bibinfo {author} {\bibfnamefont {S.}~\bibnamefont {Massar}},
  \bibinfo {author} {\bibfnamefont {G.}~\bibnamefont {Simion}}, \bibinfo
  {author} {\bibfnamefont {D.}~\bibnamefont {Wan}}, \bibinfo {author}
  {\bibfnamefont {K.}~\bibnamefont {De~Greve}},\ and\ \bibinfo {author}
  {\bibfnamefont {W.}~\bibnamefont {Wernsdorfer}},\ }\bibfield  {title}
  {\bibinfo {title} {Industrial 300 mm wafer processed spin qubits in natural
  silicon/silicon-germanium},\ }\href
  {https://doi.org/10.1038/s41534-025-01016-x} {\bibfield  {journal} {\bibinfo
  {journal} {npj Quantum Inf.}\ }\textbf {\bibinfo {volume} {11}},\ \bibinfo
  {pages} {59} (\bibinfo {year} {2025})}\BibitemShut {NoStop}%
\bibitem [{\citenamefont {Thomas}\ \emph {et~al.}(2025)\citenamefont {Thomas},
  \citenamefont {Ciriano-Tejel}, \citenamefont {Wise}, \citenamefont {Prete},
  \citenamefont {de~Kruijf}, \citenamefont {Ibberson}, \citenamefont {Noah},
  \citenamefont {Gomez-Saiz}, \citenamefont {Gonzalez-Zalba}, \citenamefont
  {Johnson},\ and\ \citenamefont {Morton}}]{Thomas2025}%
  \BibitemOpen
  \bibfield  {author} {\bibinfo {author} {\bibfnamefont {E.~J.}\ \bibnamefont
  {Thomas}}, \bibinfo {author} {\bibfnamefont {V.~N.}\ \bibnamefont
  {Ciriano-Tejel}}, \bibinfo {author} {\bibfnamefont {D.~F.}\ \bibnamefont
  {Wise}}, \bibinfo {author} {\bibfnamefont {D.}~\bibnamefont {Prete}},
  \bibinfo {author} {\bibfnamefont {M.}~\bibnamefont {de~Kruijf}}, \bibinfo
  {author} {\bibfnamefont {D.~J.}\ \bibnamefont {Ibberson}}, \bibinfo {author}
  {\bibfnamefont {G.~M.}\ \bibnamefont {Noah}}, \bibinfo {author}
  {\bibfnamefont {A.}~\bibnamefont {Gomez-Saiz}}, \bibinfo {author}
  {\bibfnamefont {M.~F.}\ \bibnamefont {Gonzalez-Zalba}}, \bibinfo {author}
  {\bibfnamefont {M.~A.~I.}\ \bibnamefont {Johnson}},\ and\ \bibinfo {author}
  {\bibfnamefont {J.~J.~L.}\ \bibnamefont {Morton}},\ }\bibfield  {title}
  {\bibinfo {title} {Rapid cryogenic characterization of 1,024 integrated
  silicon quantum dot devices},\ }\href
  {http://dx.doi.org/10.1038/s41928-024-01304-y} {\bibfield  {journal}
  {\bibinfo  {journal} {Nat. Electron.}\ }\textbf {\bibinfo {volume} {8}},\
  \bibinfo {pages} {75} (\bibinfo {year} {2025})}\BibitemShut {NoStop}%
\bibitem [{\citenamefont {Doherty}\ and\ \citenamefont
  {Wardrop}(2013)}]{doherty_two-qubit_2013}%
  \BibitemOpen
  \bibfield  {author} {\bibinfo {author} {\bibfnamefont {A.~C.}\ \bibnamefont
  {Doherty}}\ and\ \bibinfo {author} {\bibfnamefont {M.~P.}\ \bibnamefont
  {Wardrop}},\ }\bibfield  {title} {\bibinfo {title} {Two-{Qubit} {Gates} for
  {Resonant} {Exchange} {Qubits}},\ }\href
  {https://doi.org/10.1103/PhysRevLett.111.050503} {\bibfield  {journal}
  {\bibinfo  {journal} {Phys. Rev. Lett.}\ }\textbf {\bibinfo {volume} {111}},\
  \bibinfo {pages} {050503} (\bibinfo {year} {2013})}\BibitemShut {NoStop}%
\bibitem [{\citenamefont {Pedersen}\ \emph {et~al.}(2007)\citenamefont
  {Pedersen}, \citenamefont {Møller},\ and\ \citenamefont
  {Mølmer}}]{Pedersen2007}%
  \BibitemOpen
  \bibfield  {author} {\bibinfo {author} {\bibfnamefont {L.~H.}\ \bibnamefont
  {Pedersen}}, \bibinfo {author} {\bibfnamefont {N.~M.}\ \bibnamefont
  {Møller}},\ and\ \bibinfo {author} {\bibfnamefont {K.}~\bibnamefont
  {Mølmer}},\ }\bibfield  {title} {\bibinfo {title} {Fidelity of quantum
  operations},\ }\href {https://doi.org/10.1016/j.physleta.2007.02.069}
  {\bibfield  {journal} {\bibinfo  {journal} {Phys. Lett. A}\ }\textbf
  {\bibinfo {volume} {367}},\ \bibinfo {pages} {47} (\bibinfo {year}
  {2007})}\BibitemShut {NoStop}%
\bibitem [{\citenamefont {Setiawan}\ \emph {et~al.}(2014)\citenamefont
  {Setiawan}, \citenamefont {Hui}, \citenamefont {Kestner}, \citenamefont
  {Wang},\ and\ \citenamefont {Das~Sarma}}]{Setiawan2014}%
  \BibitemOpen
  \bibfield  {author} {\bibinfo {author} {\bibfnamefont {F.}~\bibnamefont
  {Setiawan}}, \bibinfo {author} {\bibfnamefont {H.-Y.}\ \bibnamefont {Hui}},
  \bibinfo {author} {\bibfnamefont {J.~P.}\ \bibnamefont {Kestner}}, \bibinfo
  {author} {\bibfnamefont {X.}~\bibnamefont {Wang}},\ and\ \bibinfo {author}
  {\bibfnamefont {S.}~\bibnamefont {Das~Sarma}},\ }\bibfield  {title} {\bibinfo
  {title} {Robust two-qubit gates for exchange-coupled qubits},\ }\href
  {https://link.aps.org/doi/10.1103/PhysRevB.89.085314} {\bibfield  {journal}
  {\bibinfo  {journal} {Phys. Rev. B}\ }\textbf {\bibinfo {volume} {89}},\
  \bibinfo {pages} {085314} (\bibinfo {year} {2014})}\BibitemShut {NoStop}%
\bibitem [{\citenamefont {Polat}\ \emph {et~al.}(2026)\citenamefont {Polat},
  \citenamefont {Overwater}, \citenamefont {Rimbach-Russ},\ and\ \citenamefont
  {Sebastiano}}]{Polat2026}%
  \BibitemOpen
  \bibfield  {author} {\bibinfo {author} {\bibfnamefont {{\.I}.}~\bibnamefont
  {Polat}}, \bibinfo {author} {\bibfnamefont {R.~W.~J.}\ \bibnamefont
  {Overwater}}, \bibinfo {author} {\bibfnamefont {M.}~\bibnamefont
  {Rimbach-Russ}},\ and\ \bibinfo {author} {\bibfnamefont {F.}~\bibnamefont
  {Sebastiano}},\ }\bibfield  {title} {\bibinfo {title} {Pulse shaping for
  ultra-fast adiabatic quantum gates},\ }\href
  {https://doi.org/10.1038/s41534-026-01245-8} {\bibfield  {journal} {\bibinfo
  {journal} {npj Quantum Inf.}\ }\textbf {\bibinfo {volume} {12}},\ \bibinfo
  {pages} {130} (\bibinfo {year} {2026})}\BibitemShut {NoStop}%
\bibitem [{\citenamefont {Kanaar}\ \emph {et~al.}(2021)\citenamefont {Kanaar},
  \citenamefont {Wolin}, \citenamefont {G{\"u}ng{\"o}rd{\"u}},\ and\
  \citenamefont {Kestner}}]{Kanaar2021}%
  \BibitemOpen
  \bibfield  {author} {\bibinfo {author} {\bibfnamefont {D.~W.}\ \bibnamefont
  {Kanaar}}, \bibinfo {author} {\bibfnamefont {S.}~\bibnamefont {Wolin}},
  \bibinfo {author} {\bibfnamefont {U.}~\bibnamefont {G{\"u}ng{\"o}rd{\"u}}},\
  and\ \bibinfo {author} {\bibfnamefont {J.~P.}\ \bibnamefont {Kestner}},\
  }\bibfield  {title} {\bibinfo {title} {Single-tone pulse sequences and robust
  two-tone shaped pulses for three silicon spin qubits with always-on
  exchange},\ }\href {https://doi.org/10.1103/PhysRevB.103.235314} {\bibfield
  {journal} {\bibinfo  {journal} {Phys. Rev. B}\ }\textbf {\bibinfo {volume}
  {103}},\ \bibinfo {pages} {235314} (\bibinfo {year} {2021})}\BibitemShut
  {NoStop}%
\bibitem [{\citenamefont {Rimbach-Russ}\ \emph {et~al.}(2023)\citenamefont
  {Rimbach-Russ}, \citenamefont {Philips}, \citenamefont {Xue},\ and\
  \citenamefont {Vandersypen}}]{RimbachRuss2023}%
  \BibitemOpen
  \bibfield  {author} {\bibinfo {author} {\bibfnamefont {M.}~\bibnamefont
  {Rimbach-Russ}}, \bibinfo {author} {\bibfnamefont {S.~G.~J.}\ \bibnamefont
  {Philips}}, \bibinfo {author} {\bibfnamefont {X.}~\bibnamefont {Xue}},\ and\
  \bibinfo {author} {\bibfnamefont {L.~M.~K.}\ \bibnamefont {Vandersypen}},\
  }\bibfield  {title} {\bibinfo {title} {Simple framework for systematic
  high-fidelity gate operations},\ }\href
  {https://doi.org/10.1088/2058-9565/acf786} {\bibfield  {journal} {\bibinfo
  {journal} {Quantum Science and Technology}\ }\textbf {\bibinfo {volume}
  {8}},\ \bibinfo {pages} {045025} (\bibinfo {year} {2023})}\BibitemShut
  {NoStop}%
\end{thebibliography}%

\end{document}